\documentclass[12pt]{iopart}
\expandafter\let\csname equation*\endcsname\relax
 \expandafter\let\csname endequation*\endcsname\relax  
\usepackage{amsmath}
\usepackage{amssymb}

\usepackage{graphicx}

\begin{document}

\title[]{Random and quasi-coherent aspects in particle motion and their
effects on transport and turbulence evolution}

\author{ M. Vlad, F. Spineanu}

\address{National Institute of Laser, Plasma and Radiation Physics \\
Magurele, Bucharest 077125, Romania}
\ead{madalina.vlad@inflpr.ro}
\vspace{10pt}
%%\begin{indented}
%%%\item[]October 2016
%%\end{indented}

\begin{abstract}
The quasi-coherent effects in two-dimensional incompressible turbulence are analyzed starting from the test particle trajectories. They can acquire coherent aspects when the stochastic potential has slow time variation and the motion is not strongly perturbed. The trajectories are, in these conditions, random sequences of large jumps and trapping or eddying events.  Trapping determines quasi-coherent trajectory structures, which have a micro-confinement effect that is reflected in the transport coefficients. They determine non-Gaussian statistics and flows associated to an average velocity. Trajectory structures also influence the test modes on turbulent plasmas. Nonlinear damping and generation of zonal flow modes is found in drift turbulence in uniform magnetic field. The coupling of test particle and test mode studies permitted to evaluate the self-consistent evolution of the drift turbulence in an iterated approach. The results show an important nonlinear effect of ion diffusion, which can prevent the transition to the nonlinear regime at small drive of the instability. At larger drive, quasi-coherent trajectory structures appear and they have complex effects on turbulence. 
\end{abstract}

\section{Introduction}

The direct numerical simulations, which have obtained important results in
the last decades, largely dominate the actual research in turbulence. The
analytical advance based on first principle description and mathematically
justifiable approximations is comparatively very small. One cause, perhaps
the principal one, is the stochastic advection process, which appears in all
turbulent systems. In realistic descriptions of turbulence the advected
field are active in the sense that they influence the velocity fields. But,
even the most simplified models that deal with passive fields have not been
solved analytically in all cases. The cause is the basic problem of particle
trajectories in stochastic velocity fields, which can be rather complex in
important special cases. One example is the turbulence in incompressible
media that is dominantly two dimensional.

The two-dimensional turbulence has a self-organizing character \cite{R1}-%
\cite{KM} that consists of the generation of quasi-coherent structure
(vortices), which can increase up to the size of the system in special cases
as the decaying turbulence in ideal fluids. The tendency to
self-organization is partly maintained is weakly perturbed systems \cite{Tzf}%
-\cite{Diamond}. This property appear at the basic level of particle
(tracer) trajectories. They are random sequences of trapping or eddying
events and long jumps. The trapping process \cite{K70}-\cite{mccomb}
strongly modifies the diffusion coefficients and leads to non-Gaussian
distribution of displacements.

Theoretical methods for determining the statistics of the advected particles
were developed in the last decades and used for the study of various aspects
of the transport. The decorrelation trajectory method (DTM, \cite{V98}) and
the nested subensemble approach (NSA, \cite{VS04}) are the first
semi-analytical methods that are able to go beyond the quasi-linear regime
that corresponds to quasi-Gaussian transport. They were specifically
developed for the case of two-dimensional incompressible turbulence that is
characterized by trajectory trapping or eddying. The general conclusion of
these studies is that the existence of space correlations of the stochastic
velocity can generate trajectory trapping or eddying, which leads to
nonstandard statistics of trajectories: non-Gaussian distributions, long
time Lagrangian correlations (memory), strongly modified transport
coefficients and an increased degree of coherence. The trapped trajectories
form quasi-coherent structures similar to fluid vortices.

Very recent results \cite{Vlad2013} have shown that these methods could be
the basis for the development of a theoretical approach for the study of
test modes in turbulent plasmas. It is similar to the Lagrangian approach
initiated by Dupree \cite{D66}, \cite{D72} and developed in the 70`s. The
assumption of random trajectories with Gaussian distribution limited the
application of Dupree's method to the quasilinear regime. The DTM and NSA
enable the evaluation of the average propagator of the test modes in the
nonlinear regime and extend the theoretical procedure to the complex
processes that are generated in these conditions. The tendency of drift
turbulence evolution beyond the quasilinear regime was obtained.

We present a theoretical approach to the study of turbulence evolution, the
iterated self-consistent method (ISC). It is based on the analysis of the
test particles and test modes in turbulent plasmas. Both test particle and
test mode studies start from given statistical description of the
turbulence. However, the coupling of these problems can lead to the
evaluation of turbulence evolution and to the understanding of the nonlinear
processes that are generated. The ISC is applied here to the drift
turbulence in magnetically confined plasmas.

The paper is organized as follows. The physical processes analyzed in this
paper and the main ideas that are followed in this study are presented in
Section 2. Section 3 contains the test particle studies of transport, the
DTM and the NSA. A discussion on the representation of the complex
trajectories by these methods is also included. The two modules of the ISC
(test particle and test mode studies) for the case of drift turbulence are
presented in Section 4. The ISC is discussed in Section 5. Section 6
contains the results on the evolution of drift turbulence and the analysis
of the nonlinear processes that appear in different conditions (weak and
strong drive of the instability). The conclusions are summarized in Section
7.

\section{Random and coherent aspects of the trajectories}

Numerical simulations have shown that particle trajectories in
two-dimensional incompressible stochastic velocity fields are characterized
by a mixture of random and quasi-coherent aspects, which appear as a random
sequence of large jumps and trapping or eddying events. They are solutions
of the equation%
\begin{equation}
\frac{dx_{i}}{dt}=-\varepsilon _{ij}\partial _{j}\phi (\mathbf{x}(t)\mathbf{,%
}t),  \label{eq1}
\end{equation}%
where $\mathbf{x}(t)$ is the trajectory in rectangular coordinates $\mathbf{x%
}=(x_{1},x_{2}),$ $\varepsilon _{12}=1,$\ $\varepsilon _{21}=-1,$\ $%
\varepsilon _{ii}=0,$\ and $\phi (\mathbf{x,}t)$\ is the potential.\ This is
a Hamiltonian system with the conjugate variables represented by the two
components of particle trajectories.

The origin of eddying is the Hamiltonian structure of Eq. (\ref{eq1}). The
trajectories are periodic and they wind on the contour lines of the
potential when it is time independent. This defines the state of permanent
trapping. The Lagrangian potential is invariant in these conditions and the
transport is subdiffusive because the particles are tied on the fixed
contour lines of the potential.

If the motion is weakly perturbed (by time variation of the potential or by
other components of the motion that can by introduced in Eq. (\ref{eq1})),
the trapping is temporary and alternates with large jumps. The statistical
importance of the trapping events depends on the strength of the
perturbation that is represented by dimensionless factors, which are usually
ratios of characteristic times. Two characteristic times are defined in (\ref%
{eq1}): the correlation time of the potential $\tau _{c}$ that gives the
scale of the time variation of $\phi (\mathbf{x,}t)$ and the time of flight
of the particles $\tau _{fl}=\lambda _{c}/V,$ where $\lambda _{c}$\ is the
space scale of $\phi (\mathbf{x,}t)$\ and $V$ is the amplitude of the
velocity. These parameters appear in the Eulerian correlation of the
potential defined in Section 3. The characteristic times define the Kubo
number $K=\tau _{c}/\tau _{fl},$\ which is a measure of trapping in time
dependent potentials. Namely, the permanent trapping corresponds to $%
K\rightarrow \infty ,$\ temporary trapping exists for $K>1,$ and the
statistical relevance of trapping is a decreasing function of $K.$ Any other
perturbation of the basic Hamiltonian motion \ defines a characteristic time 
$\tau _{d}$ for the decorrelation of the trajectory from the potential, and
a dimensionless parameter similar to Kubo number, the decorrelation number $%
K_{d}=\tau _{d}/\tau _{fl}.$\ 

The trapping events appear when the trajectory arrives around the maxima and
the minima of the potential. The process is quasi-coherent because it
affects all the particles situated in these regions in the same way. The
high degree of coherence of the trapped trajectories is evidenced in a study
of the statistical properties of the distance $\delta x$ between neighbour
trajectories \cite{VS04}. The time evolution of $\left\langle \delta
x^{2}(t)\right\rangle $ is very slow, which shows that neighbour particles
have a coherent motion for a long time compared to $\tau _{fl}.$ They are
characterized by a strong clump effect with the increase of $\left\langle
\delta x^{2}(t)\right\rangle $ that is slower than the Richardson law. These
trajectories form intermittent quasi-coherent structures similar to fluid
vortices and represent eddying regions. Their average size and life-time
depend on the characteristic of the turbulence.

The trapping events appear around the maxima and the minima of the
potential. The other particles (that evolve on contour lines corresponding
to small absolute values of the potential) have much larger displacements.
The large displacements are caused by the large size of these contour lines
and also by the decorrelation, which allows transitions between neighbour
potential cells. These large jumps are random. The transport is essentially
determined by them, while the trapping events have negligible contributions.
Thus, trapping can be associated to a process of micro-confinement. At a
given moment, the micro-confinement affects a fraction of the particle, and,
in a large time interval, every particle is subjected to micro-confinement
events during a fraction of this time. The micro-confinement determines the
existence of transport reservoirs in the sense that diffusion can be
strongly enhanced when these trapped particles are released by the increase
of the perturbation strength.

The quasi-coherent trajectory structures have more complex effects in the
presence of an average velocity $V_{d}.$ The probability of the Lagrangian
velocity has to be time invariant, as required by the zero divergence of the
velocity field. In particular, the average Lagrangian velocity $V(t)$ has to
be equal to the average Eulerian velocity at any time. The trapped particles
do not contribute to $V(t),$ which means that the free particles have an
average velocity larger than $V_{d}$ in order to compensate the fraction of
trapped particles $n_{tr}.$ This effect leads, in the case of a potential
that drifts with the velocity $V_{d},$ to flows in both directions: the
structures move with the potential with the velocity $V_{d}$ and the free\
particles have an average motion in the opposite direction with a velocity $%
V_{f}$ such that\ \ 
\begin{equation}
n_{tr}V_{d}+n_{f}V_{f}=0.  \label{flow}
\end{equation}%
The average velocity also determines the release of a part of the trapped
particles and the decrease of the size of the trajectory structures. At
large $V_{d}$, the structures are completely eliminated at 

\bigskip

The aim of the paper is to analyze the effects produced by these complex
trajectories on the evolution of drift turbulence and on the associated
transport.

We show that the physical picture of the trajectories that mix random and
quasi-coherent aspects is well represented in the DTM and NSA. The problem
of test particle transport and typical results obtained with these methods
are presented in Section 3. The image of the random and quasi-coherent
aspects given by the DTM and NSA is emphasized and the effects of the
micro-confinement are discussed in this Section.

The drift turbulence is analyzed in relation with the random and
quasi-coherent aspects of the trajectories (Sections 4 and 6). This type of
turbulence determines special effects on transport coefficients, which are
presented in Section 4.2, together with the characteristics of the
trajectory structures and their influence on the probability of
displacements. The test modes on turbulent plasma are presented in Section
4.1. It is shown that the trajectory structures can lead to strong nonlinear
effects. The self-consistent evolution of the drift turbulence obtained with
the ISC is finally analyzed in Section 6.

\section{Test particle approach to stochastic transport}

The Eulerian and the Lagragian correlations are the main concepts in test
particle transport. The two point Eulerian correlation (EC) of the velocity
is defined as%
\begin{equation}
E_{ij}(\mathbf{x},t)\equiv \left\langle v_{i}(\mathbf{x}_{1},t_{1})~v_{j}(%
\mathbf{x}_{2},t_{2})\right\rangle ,  \label{ECdef}
\end{equation}%
where $\mathbf{x=x}_{2}-\mathbf{x}_{1},$ $t=t_{2}-t_{1}$ and $i,j=1,2$
account for the two components of the velocity. The main statistical
parameters of the velocity field appear in this function. The amplitudes of
velocity fluctuation are $V_{i}^{2}=E_{ii}(\mathbf{0},0),$ the correlation
lengths $\lambda _{i}$\ are the characteristic decay lengths of the
functions $E_{ii}(\mathbf{x},0),$ and the correlation time is defined by the
time decay of $E_{ij}(\mathbf{0},t).$ We consider here homogeneous and
stationary velocity fields, which have ECs that depend only on the distance
between the two points and on the difference between the two times, as (\ref%
{ECdef}).

The EC describes the space structure and the time variation of the
stochastic velocity field

The Lagrangian velocity correlation (LVC) is defined as%
\begin{equation}
L_{ij}(t)\equiv \left\langle v_{i}(\mathbf{x}_{1},t_{1})~v_{j}(\mathbf{x}%
(t;t_{1},\mathbf{x}_{1})\right\rangle ,  \label{LCdef}
\end{equation}%
where $\mathbf{x}(t;t_{1},\mathbf{x}_{1})$\ is the trajectory in a
realization of the stochastic field that starts at $t=t_{1}$ in the point $%
\mathbf{x}_{1}$.

The LVC is thus a time dependent function that describes statistical
properties of particle motion in the stochastic velocity field. In most
cases the LVC decays to zero since the Lagrangian velocity becomes
statistically independent on the initial velocity (it decorrelates). The
characteristic time $\tau _{d}$\ of the LVC decay is the measure of the
memory of the stochastic Lagragian velocity. The decorrelation time $\tau
_{d}$ is the correlation time $\tau _{c}$\ of the EC if particle motion is
determined only by the\ velocity field, and it can be completely different
of $\tau _{c}$ in the presence of additional components of the motion
(average velocity, collisional diffusivity, etc.)

Besides this important effect of memory, the LVC determines the main
quantities related to test particle transport. As shown by Taylor \cite%
{Taylor}, the mean square displacement (MSD) $\left\langle
x_{i}^{2}(t)\right\rangle $ and its derivative, the running diffusion
coefficient $D_{i}(t)$, are integrals of the LVC%
\begin{equation}
\left\langle x_{i}^{2}(t)\right\rangle =2\int_{0}^{t}d\tau \;L_{ii}(\tau
)\;(t-\tau ),  \label{MSD}
\end{equation}%
\begin{equation}
D_{ii}(t)=\int_{0}^{t}d\tau \;L_{ii}(\tau ).  \label{D}
\end{equation}%
\ \ \ \ \ The asymptotic values of $D_{i}(t)$\ 
\begin{equation}
D_{i}=\underset{t\rightarrow \infty }{\lim }D_{ii}(t)=\int_{0}^{\infty
}L_{ii}(\tau )d\tau  \label{das}
\end{equation}%
are the diffusion coefficients. The time dependent diffusion coefficients
provide\ the "microscopic" characteristics of the transport process. The
diffusion at the transport space-time scales, which are much larger than $%
\lambda _{c}$ and $\tau _{c},$ is described by the asymptotic values $D_{i}.$

The aim of the test particle studies is essentially to determine the LVC as
function of the EC.

The decorrelation of the Lagrangian velocity after the memory time $\tau
_{c} $\ leads to the possibility of decomposing the statistics of the
displacements at large time $t>>\tau _{c}$\ in a sequence of independent
small scale processes of time $\tau _{c}.$\ The probability of the small
scale displacements (the micro-probability), $P^{\tau _{c}}(\mathbf{x},t),$
determines the elementary step of the transport process as the MSD%
\begin{equation*}
\Delta _{i}^{2}\equiv \left\langle (x_{i}(t_{1}+\tau _{c};t_{1},\mathbf{x}%
_{1})-\mathbf{x}_{1})^{2}\right\rangle =\int d\mathbf{x}~x_{i}^{2}P^{\tau
_{c}}(\mathbf{x},\tau _{c}).
\end{equation*}%
The statistics of the displacements at large times $t>>\tau _{c}$ is
Gaussian, with the exception of the processes that have $\Delta \rightarrow
\infty $\ or $\tau _{c}\rightarrow \infty $\ (see\ \cite{Bouchaud}, \cite%
{Bcarte}).

\bigskip

The asymptotic diffusion coefficients are usually approximated by a random
walk with the step $\Delta $\ performed in the time $\tau _{c},$\ \bigskip
that is $D_{rw}\cong \Delta ^{2}/\tau _{c}.$

It is important to underline that the test particle transport gives results
that are in very good agreement to those obtained from turbulent fluxes when
the space scale of the turbulence is much smaller than the characteristic
lengths of the gradients. This is confirmed by the numerical simulations 
\cite{Naulin}, \cite{Jenko}.

\bigskip

Most of the theoretical methods are essentially based on Corrsin hypothesis 
\cite{Corrsin1959}, which assume that the micro-probability $P^{\tau _{c}}(%
\mathbf{x},t)$ is Gaussian and that the displacements are statistically
independent on the velocity field. They lead to diffusive transport in the
limit of static velocity fields ($\tau _{c}\rightarrow \infty ),$ and thus
they cannot apply to the special case of zero divergence two-dimensional
velocity fields. It can be shown that a diffusive transport is obtained even
when the second assumption is eliminated, which suggest that the
micro-probability is not Gaussian in the presence of trapping. The first
theoretical approach that finds subdiffusive transport in this special case (%
\cite{Gruzinov}, \cite{Isichenko}) does not use the Gaussian assumption, but
it is based on the percolation theory. It only determines the scaling of the
asymptotic diffusion coefficients. The detailed statistical information that
is contained in the LVC (\ref{LCdef}) was first obtained by the
decorrelation trajectory method (DTM, \cite{V98}). This semi-analytical
method was developed and validated by the nested subensemble approach (NSA, 
\cite{VS04}).

A short review of the DTM and NSA and a discussion of the image they give on
the micro-confinement process is presented below.

\subsection{The DTM and NSA}

Trajectory trapping or eddying is a consequence of the Lagrangian invariance
of the potential. The DTM and the NSA were developed having in mind the
necessity of evidencing the statistical consequences of this invariance. We
have also imposed the condition of performing only the approximations that
do not violate this property.

The main idea is to define subensembles of the realizations of the potential
that correspond to the same values of the potential and of its derivatives
in the starting point of the trajectories $\mathbf{x=0,}$ $t=0.$\ A system
of nested subensembles is constructed. All the realization with the
potential $\phi (\mathbf{0},0)=\phi ^{0}$ are grouped together in a
subensemble $S_{0},$ then $S_{0}$ is subdivided\ into subensembles $S_{1}$\
according to the values of the first space derivatives of the potential $%
\phi _{i}^{0}$, then each $S_{1}$\ is again divided\ into smaller
subensembles $S_{2}$\ using the values of the second derivatives $\phi
_{ij}^{0}$, and the procedure can continue. The trajectories in a
subensemble $S_{n}$ are super-determined, in the sense that they have
supplementary initial condition (the potential $\phi ^{0},$ the initial
velocity that is the derivative of the potential $\mathbf{v}(\mathbf{0},0)=%
\mathbf{v}^{0}=\left( -\phi _{2}^{0},~\phi _{1}^{0}\right) $, and higher
order derivatives up to the order $n).$ This leads to a high degree of
similarity of the trajectories in a subensemble $S_{n}$, that increases with
the increase of the nesting order $n.$

The statistical description of the velocity in a subensemble $S_{n}$ can be
derived from the statistics in the whole set of realization using
conditional averages. A Gaussian velocity reduced at the subensemble $S_{n}$
remains Gaussian, but its average and dispersion are modified. The
subensemble average $\mathbf{V}^{S_{n}}(\mathbf{x},t)\equiv \left\langle 
\mathbf{v}(\mathbf{x},t)\right\rangle _{S_{n}}$ exists even in the case of
zero average velocities. It is a function of the EC that also depends
(linearly) on the set of parameters of the nested subensembles. The
subensemble average velocity $\mathbf{V}^{S_{n}}(\mathbf{x},t)$ is a
space-time dependent function, which has the value $\mathbf{V}^{S_{n}}(%
\mathbf{0},0)=\mathbf{v}^{0}$ (where $\mathbf{v}^{0}=\left( -\phi _{2}^{0},\
\phi _{1}^{0}\right) ),$ \ as imposed by the conditions that define $S_{1}$\
in the nested subensembles. The subensemble dispersion $\left\langle \left(
\delta \mathbf{v}(\mathbf{x},t)\right) ^{2}\right\rangle _{S_{n}}$ (where $%
\delta \mathbf{v}(\mathbf{x},t)=\delta \mathbf{v}(\mathbf{x},t)-\mathbf{V}%
^{S_{n}}$) is much smaller that the amplitude of the velocity fluctuations
in the whole set of realization $\mathbf{V}^{2}$. It is zero for $\mathbf{x=0%
}$ and $t=0$\ (because imposing a condition on the value of the velocity
eliminates any fluctuation), and it reaches the value $\mathbf{V}^{2}$ at
distances larger than the correlation lengths $\lambda _{i}$ and/or at times
larger than the correlation time $\tau _{c}$. The small amplitude of the
velocity fluctuations in the subensemble $S_{n}$ provides a second
contribution to the increase of the degree of similarity of the set of
trajectories that yield from the realizations that are included in $S_{n}.$

The subensemble average velocity $\mathbf{V}^{S_{n}}(\mathbf{x},t)$ has the
same structure as particle velocity. It can be derived from a function $\Phi
^{S_{n}}(\mathbf{x},t),$ which is the subensemble average potential%
\begin{equation}
V_{i}^{S_{n}}(\mathbf{x},t)=-\varepsilon _{ij}\frac{\partial \Phi ^{S_{n}}(%
\mathbf{x},t)}{\partial _{x_{j}}}.  \label{V-Fi}
\end{equation}%
The potential $\Phi ^{S_{n}}(\mathbf{x},t)$ is obtained as the conditional
average with the set of conditions that define the nested subensembles.
Thus, it is equal to $\phi ^{0}$\ for $\mathbf{x=0}$ and $t=0$\ for any
value of the nesting order $n.$ \ \ 

The existence of the subensemble average velocity $\mathbf{V}^{S_{n}}(%
\mathbf{x},t)$ determines an average trajectory in $S_{n}$ that is obtained
by performing the subensemble average of the equation of motion%
\begin{equation}
\frac{d}{dt}\left\langle \mathbf{x}(t)\right\rangle _{S_{n}}=\left\langle 
\mathbf{v}(\mathbf{x}(t),t)\right\rangle _{S_{n}}.  \label{avec}
\end{equation}%
\ The right hand side of this equation is generally very difficult to be
evaluated since it is the average of a stochastic function $\mathbf{v}(%
\mathbf{x},t)$ of a stochastic variable, the trajectory $\mathbf{x}(t)$ that
is determined by $\mathbf{v}(\mathbf{x},t)$. However, the high degree of
similarity of the trajectories in the subensemble $S_{n}$ enables an
important simplification that consists in neglecting the fluctuations of the
trajectories. The average Lagrangian velocity is thus approximated in $S_{n}$
by the subensemble Eulerian velocity calculated along the average trajectory%
\begin{equation}
\frac{d}{dt}\mathbf{X}^{S_{n}}(t)=\mathbf{V}^{S_{n}}(\mathbf{X}%
^{S_{n}}(t),t),  \label{DT}
\end{equation}%
where $\mathbf{X}^{S_{n}}(t)$ is the approximate subensemble average
trajectory in $S_{n}$, $\mathbf{X}^{S_{n}}(t)\cong \left\langle \mathbf{x}%
(t)\right\rangle _{S_{n}}.$ It is called the decorrelation trajectory (DT),
because it shows the way toward decorrelation in each subensemble. We note
that the approximation (\ref{DT}) of Eq. (\ref{avec}) is in agreement with
the invariance of the Lagrangian potential. Equation (\ref{DT}) has
Hamiltonian structure as the equation of particle motion (\ref{eq1}). This
leads, in the case of static velocity fields, to the invariance of the
Lagrangian potential $\Phi ^{S_{n}}(\mathbf{X}^{S_{n}}(t)),$ which is equal
to $\phi ^{0}$\ for any value of the nesting order $n.$

The LVC, $D_{ii}(t)$ and other Lagrangian averages are determined by summing
the contribution of all subensembles. This leads to integrals over the
parameters that define the subensembles.

Thus the NSM is a systematic expansion based on the nested subensembles. It
is a semi-analytical method that determines the statistics of the stochastic
trajectories in terms of the DTs. These trajectories and the integrals over
the subensembles have to be numerically calculated. The calculations are at
PC level, with run times that increase as the nesting order increases,
starting from few minutes for $n=2$.

The DTM \cite{V98} corresponds to the nesting order $n=1.$ The results
obtained using the NSM of order $n=2$ are presented in \cite{VS04}. The DTs
for the subensembles $S_{2}$\ are completely different of those of the DTM.
However they lead to results for the LVC that are\ not much different of
those obtained with the DTM, which shows that the NSM converges fast. The
higher order NSM provides more detailed statistical information at the
expense of the increase of the number of the DTs and of the complexity of
their equation. The statistics of the distance between neighbor trajectories
was determined with the $n=2$ NSM.

We use here the DTM that is based on the subensembles $S_{0}$\ and $S_{1}.$\
They are defined by the conditions $\phi (\mathbf{0},0)=\phi ^{0}$\ and $%
\mathbf{v}(\mathbf{0},0)=\mathbf{v}^{0}$. The subensemble $S_{1}$ average
potential is 
\begin{equation}
\Phi ^{S_{1}}(\mathbf{x},t)=\phi ^{0}\frac{E(\mathbf{x},t)}{E(0)}+v_{1}^{0}%
\frac{\partial _{2}E(\mathbf{x},t)}{V_{1}^{2}}-v_{2}^{0}\frac{\partial _{1}E(%
\mathbf{x},t)}{V_{2}^{2}},  \label{FS}
\end{equation}%
where $\partial _{i}E$ are the space derivatives of the EC. Summing the
contributions of all subensembles, the Lagragian velocity correlation and
the time dependent diffusion coefficient are evaluated as%
\begin{eqnarray}
L_{ij}(t) &=&\int \int d\phi ^{0}d\mathbf{v}^{0}P(\phi ^{0})P(\mathbf{v}%
^{0})\ v_{i}^{0}\ V_{j}^{S}\left[ \mathbf{X}^{S}(t),v_{th}t,t\right] ,
\label{Lvc} \\
D_{ij}(t) &=&\int \int d\phi ^{0}d\mathbf{v}^{0}P(\phi ^{0})P(\mathbf{v}%
^{0})\ v_{i}^{0}\ X_{j}^{S}(t),  \label{dij}
\end{eqnarray}%
where $P(\phi ^{0})$ and $P(\mathbf{v}^{0})$\ are the Gaussian probabilities
for $\phi ^{0}$ and $\mathbf{v}^{0}$\ respectively.

An important simplification was recently found \cite{VS2015}. We have shown
that the number of parameters of the subensembles can be reduced to only two
without significant modifications of the diffusion coefficients. The
magnitude $u$ of the normalized velocity $v_{i}(\mathbf{0},0)/V_{i}$ can be
eliminated from the definition of $S_{1}.$ The integral over $u$ can be
performed in the subensemble Eulerian correlation in $S_{1}.$\ This leads to
a modified average potential, which depends only on $\phi ^{0}$ and $\theta
^{0},$\ the orientation of the normalized velocity\ 
\begin{equation}
\Phi ^{S_{1}^{\prime }}(\mathbf{x},t)=\phi ^{0}\frac{E(\mathbf{x},t)}{E(0)}+%
\sqrt{\frac{8}{\pi }}\cos (\theta ^{0})\frac{\partial _{2}E(\mathbf{x},t)}{%
V_{1}}-\sqrt{\frac{8}{\pi }}\sin (\theta ^{0})\frac{\partial _{1}E(\mathbf{x}%
,t)}{V_{2}}.  \label{FSprim}
\end{equation}%
The parameter $\theta ^{0}$ defines a larger subensemble $S_{1}^{\prime }$
that includes the subensembles $S_{1}$ with any value of $u.$

The time dependent diffusion coefficients are obtained in this case from%
\begin{eqnarray}
D_{11}(t) &=&\frac{V_{i}}{2\pi }\sqrt{\frac{\pi }{2}}\int_{-\infty }^{\infty
}d\phi ^{0}P(\phi ^{0})\int_{0}^{2\pi }\ d\theta ^{0}\cos (\theta
^{0})X_{1}^{S^{\prime }}(t),  \label{Dprim} \\
D_{22}(t) &=&\frac{V_{i}}{2\pi }\sqrt{\frac{\pi }{2}}\int_{-\infty }^{\infty
}d\phi ^{0}P(\phi ^{0})\int_{0}^{2\pi }\ d\theta ^{0}\sin (\theta
^{0})X_{2}^{S^{\prime }}(t)
\end{eqnarray}%
where $\mathbf{X}^{S^{\prime }}(t)$ is the DT in the subensemble $S^{\prime }
$ that is the solution of%
\begin{equation}
\frac{dX_{i}^{S^{\prime }}}{dt}=V_{i}^{S^{\prime }}(\mathbf{X}^{S^{\prime
}},t)+\delta _{i2}V_{d},  \label{DTprim}
\end{equation}%
where we have introduced an average velocity $V_{d}$ along $\mathbf{e}_{2}$
axis.

Thus, the diffusion coefficients are obtained by calculating the double
integral in Eq. (\ref{Dprim}) with the DTs determined for each value of $%
\phi ^{0},$ $\theta ^{0}$ from Eq. (\ref{DTprim}), where the velocity $%
V_{i}^{S^{\prime }}$ is obtained from the potential (\ref{FSprim}).

Other Lagrangian quantities are obtained using expression similar to (\ref%
{Dprim}), which consists of using the DTs for the evaluation of the average
Lagragian quantities in $S^{\prime }$ and of the summation of the
contributions of all subensembles.

This is a very fast version of the DTM because the number of DTs is strongly
reduced (from $N^{3}$ to $N^{2},$\ where $N$ is the number of discretization
points for each parameter of the subensembles).

\subsection{Trapping and micro-confinement description by DTM}

The decorrelation trajectories (DTs) are the main concept of this
theoretical description of trajectory statistics. They are approximations of
the average trajectories in the subensembles, which embed the high degree of
similarity of the corresponding sets of stochastic trajectories. The
probability of each DT is the probability of the parameters of its nested
subensembles, which are initial conditions for the DT. Their average on the
whole set of realizations (obtained by integrals similar to (\ref{Dprim}))
is zero. The DTs are in agreement with the invariance of the Lagrangian
potential in the static case. Each DT is a periodic functions in this case,
and it is tied to the contour line of the subensemble average potential. The
Lagragian potential for the DT is $\phi ^{0},$ the same with the Lagrangian
potential for any trajectory included in $S_{0}.$

The DTs describe the Lagragian effects at the turbulence space-time scale.
After times longer than the decorrelation time $\tau _{d}$\ or after
displacements larger than the correlation lengths $\lambda _{i},$\ all DTs
saturate. The asymptotic values $X_{i}^{S^{\prime }}(t)\rightarrow l_{i}^{d}$
are functions of the parameters of the corresponding nested subensembles,
and they represent the average distances traveled during the decorrelation
time by the trajectories in the subensembles. Besides, the DTs provide a
detailed description of the decorrelation process, which consists of the
path through the correlated zone in each subensemble. In particular,
trajectory trapping corresponds to $l_{i}^{d}<\lambda _{i}$\ for a
statistically relevant number of DTs.

The DTs obtained from Eq. (\ref{DTprim}) for $V_{d}=0$ strongly depend on
the correlation time $\tau _{c}$ and on the value of $\left\vert \phi
^{0}\right\vert .$

For fast decorrelation (small $\tau _{c}),$\ the DTs are along the initial
velocity and they practically do not depend on the initial potential. This
type of DTs are specific to the Gaussian transport. The fast decorrelation
determines quasilinear (Gaussian) transport.

In the case of slow decorrelation (large $\tau _{c}),$ the DTs are
completely different for small and large values of the initial potential. At
large values of $\left\vert \phi ^{0}\right\vert $, the DTs rotate on closed
paths with a velocity that decay to zero at times larger than $\tau _{c}.$
The asymptotic displacements are smaller than the correlation length. The
DTs with small $\left\vert \phi ^{0}\right\vert $ have much larger
displacements that are predominantly oriented along the direction of the
initial velocity. Thus, the typical structure of the real trajectories (that
consists of a random sequence of eddying events and jumps) is represented by
the DTs. They actually describe segments of trajectories of time intervals
of the order $\tau _{c}$ that are either trapping events or free
displacements, depending on the initial potential. More important, the
statistical relevance of each type of motion and the influence of the
trajectory trapping is determined by the whole set of DTs. The slow
decorrelation determines nonlinear (non-Gaussian) transport.

\bigskip

The existence of an average velocity $V_{d}$ in the equation of motion
determines modifications of the DTs in the nonlinear regime. The value of $%
\left\vert \phi ^{0}\right\vert $ that gives the limit between free and
trapped trajectories increases, which means that the fraction of trapped
trajectories decreases. The average velocity also produces a new type of
DTs. They are opened trajectories that turn toward the direction of $V_{d}%
\mathbf{e}_{2}.$Their number increases as $V_{d}$\ increases and, when $%
V_{d} $\ is much large than the stochastic velocity, all DTs are of this
type. Thus, the average velocity eliminates progressively (as $V_{d}$
increases)\ the trapping of the trajectories and eventually leads to
quasilinear transport.

%%%%%%%%%

\begin{figure}[tbp]
\centerline{\includegraphics[height=6cm]{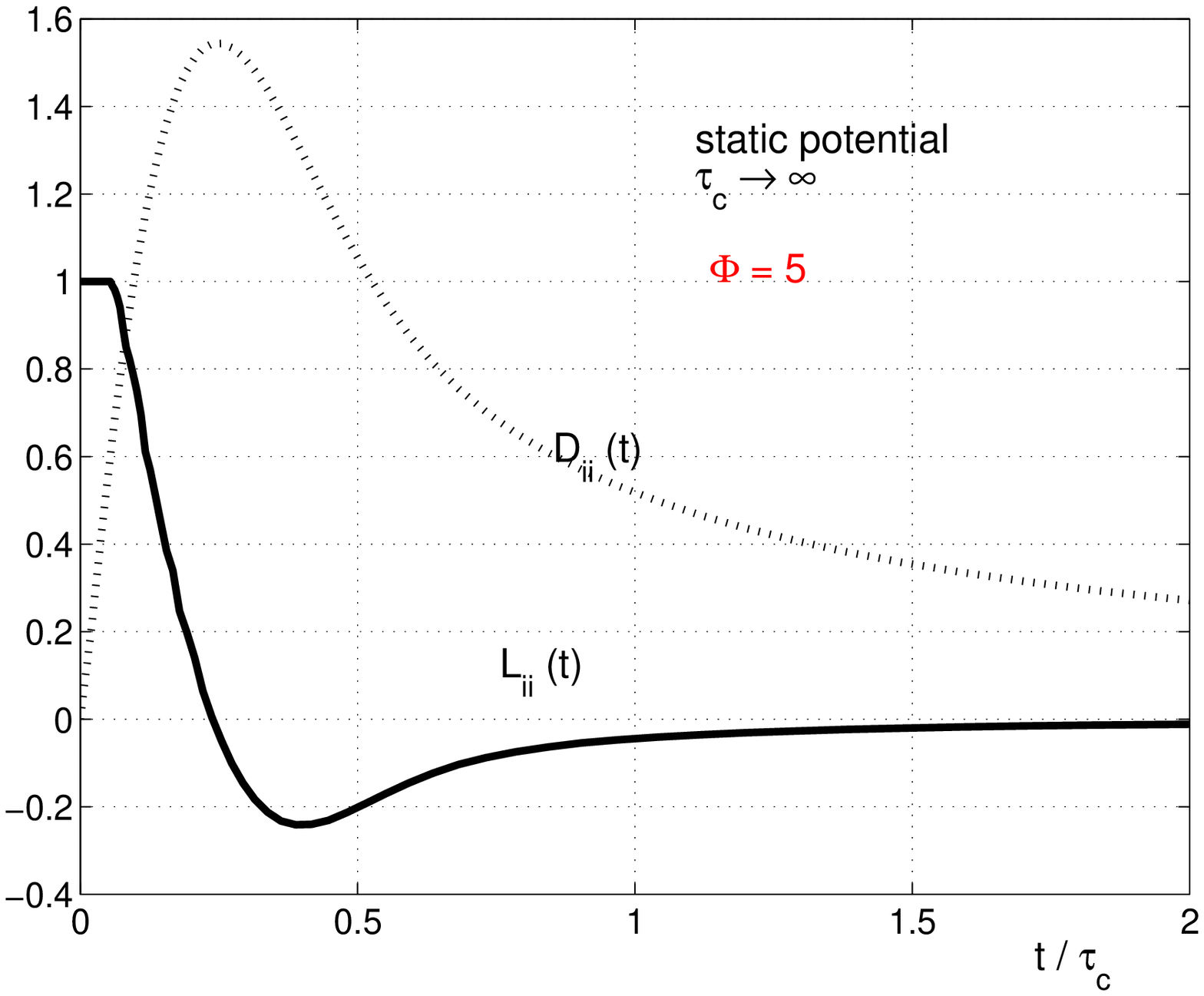}} \centerline{%
\includegraphics[height=6cm]{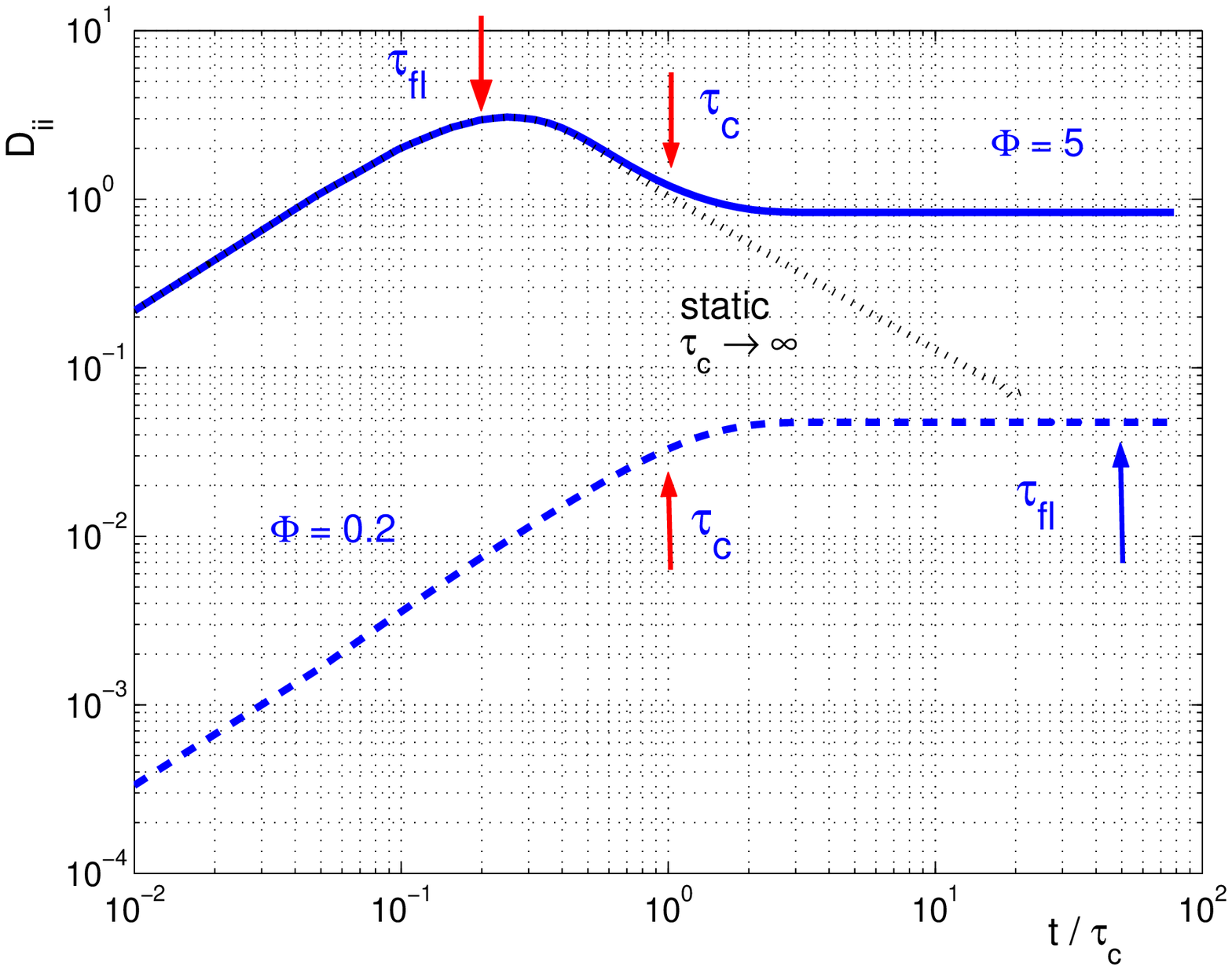}}
\caption{ a) The LVC and $D_{ii}(t)$ for the subdiffusive transport ($%
\protect\tau_c\rightarrow \infty$).\newline
b) $D_{ii}(t)$ in the quasilinear (dashed line for $\Phi=0.2$) and nonlinear
(continuous line for $\Phi=5$) regimes for the correlation time $\protect\tau%
_c=1. $ The subdiffusive transport is also represented for comparison
(dotted line). }
\label{Figure1}
\end{figure}

%%%%%%%%%

\bigskip

Typical results for $D_{ii}(t)$ and for the LVC are shown in Figure 1. They
are obtained for isotropic turbulence with the EC of the potential $E=$\ $%
\Phi ^{2}\exp (-(x_{1}^{2}+x_{2}^{2})/\lambda ^{2}/2-t/\tau _{c})$ and $%
V_{d}=0.$ The diffusion process is isotropic ($D_{11}=D_{22})$ and the
decorrelation is produced by the time variation of the potential ($\tau
_{d}=\tau _{c}).$

Figure 1a presents the case of static potentials ($\tau _{c}\rightarrow
\infty $). One can see that the diffusion coefficient has a maximum at a
time of the order of the time of flight $\tau _{fl,i}=\lambda _{i}/V_{i},$
and then it decays to zero. The DTM yields subdiffusive transport for such
fields where the trajectories are tied on the contour lines of the
potential. The LVC defined by (\ref{LCdef}) is negative at large time
(Figure 1a), such that its integral decays to zero. It has a negative tail
that decays as a negative power of time.

When the potential is not static ($\tau _{c}$ is finite), the transport
becomes diffusive. As seen in Figure 1b, $D_{ii}(t)$\ has different shapes
at small and large amplitudes of the potential fluctuations $\Phi .$\ The
diffusion coefficient increases linearly at small time and it eventually
saturates (for $t>\tau _{c})$ in the case of small $\Phi $\ (dashed line),
while, at large $\Phi $\ (continuous line),\ a transitory decay appears
before saturation. The decay accounts for the micro-confinement produced by
trajectory trapping. This process appears if $\tau _{fl}<\tau _{c}$ (in
general, for $\tau _{fl}<\tau _{d}).$ One can see in Figure 1b that the
variation of $\tau _{c}$\ determines opposite effects in the two cases. The
decrease of $\tau _{c}$ leads to the decrease of the asymptotic diffusion
coefficient in the quasilinear case and to the increase of $D_{i}$\ for the
nonlinear transport.\ \ 

\ 

%%%%%%%%%

\begin{figure}[tbp]
\centerline{\includegraphics[height=6cm]{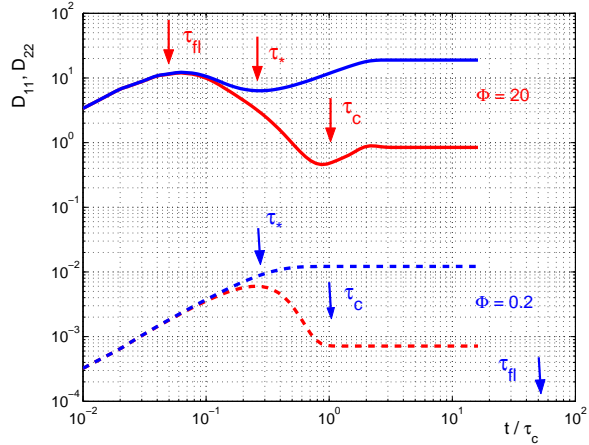}}
\caption{The anisotropic diffusion coefficients in the presence of an
average velocity $V_d{\mathbf{e}}_2$: $D_{11}(t)$ (red) and $D_{22}(t)$
(blue) in the quasilinear regime (dashed lines for $\Phi=0.2 $) and in the
nonlinear regime (continuous lines for $\Phi=20$) }
\label{Figure2}
\end{figure}

%%%%%%%%%

The effect of an average velocity $V_{d}$ is shown in Figure 2. The
diffusion becomes non-isotropic, but the differences appear in $D_{ii}(t)$
only for $t>\tau _{\ast },$\ where $\tau _{\ast }$ is the characteristic
time defined by the average velocity $\tau _{\ast }=\lambda /V_{d}.$ In the
case of the quasilinear transport (dashed lines for $\Phi =0.2),$\ both
coefficients decrease, but $D_{11}$ much more.\ It saturates for $t>\tau
_{c},\ $after a significant decrease during the interval $(\tau _{\ast
},\tau _{c}).$ The diffusion coefficient\ $D_{22}$\ saturates at $\tau
_{\ast }$\ and it does not depend on the decorrelation time if $\tau
_{d}>\tau _{\ast }.$\ This shows that the average velocity determines the
decorrelation along its direction and has a decaying effect for the
diffusion along the density gradient. In the nonlinear regime, the average
velocity determines a strong increase of $D_{22}$\ and a small decrease of $%
D_{11},$ as seen in Figure 2 (the continuous lines).\ \ \ 

\bigskip

The LVC for the unperturbed static potential (Figure 1a) is the result of
the permanent trapping. It shows that the process has a long memory (the
decay at large time of the LVC is of the power law type $L_{ii}(t)\sim
t^{-\nu },$ where $\nu >0).$ Its positive and negative parts compensate,
such that the time integral, which in the diffusion coefficient $D_{ii}(t),$%
\ decays to zero at large time. This behaviour is the representation of the
micro-confinement.

Any small perturbation (characterized by a large decorrelation time $\tau
_{d})$ has significant effect due to the existence the long time correlation
of the Lagrangian velocity. It destroys the equilibrium between the positive
and the negative parts of the LVC and leads to finite values of the
asymptotic diffusion coefficient. For instance, the time variation of the
potential\ determines the cutting of the LVC at the correlation time $\tau
_{c}$\ and consequently the saturation of $D_{ii}(t),$ as seen in Figure 1b.
This is the representation of the perturbation of the micro-confinement by
releasing a part of trajectories. The transport reservoir is the maximum of $%
D_{ii}(t)$\ (Figure 1a). It provides the maximum diffusion coefficient that
correspond the a decorrelation process with $\tau _{d}=\tau _{fl},$ which
can release all the trapped trajectories.

Various aspects of the transport in fusion plasmas were studied using the
DTM, that was developed and adapted to the study of models with increased
complexity \cite{VS2015}-\cite{VS16}.

\section{Test modes and test particles in drift turbulence}

Drift instabilities are low-frequency modes generated in nonuniform
magnetically confined plasmas \cite{GR}. We consider here the basic drift
instability in uniform magnetic field. It belong to the family of
instabilities that have the main role in particle and energy transport in
fusion research \cite{K02}, \cite{Horton}, \cite{Garbet}. They have complex
nonlinear evolution with generation of large correlations and zonal flow
modes \cite{Tzf}-\cite{Diamond}.

Test modes and test particle studies start from the given statistical
description of turbulence. They provide answers to different questions
compared to the case self-consistent studies. Namely, they evaluate the
growth rates of the modes, the diffusion coefficients and the
characteristics of the transport as functions of the parameters of the
background turbulence. The self-consistent studies determine the
characteristics of the turbulence generated in given macroscopic conditions
(as density gradient, temperatures) and the associated transport. We show in
Section 6 that an iterated self-consistent (ISC) method can be developed for
the drift turbulence. It is based on the coupled study of test particle and
test modes. These main modules of the ISC theory are presented here.

The DTM and the NSA provide the basis for the development of a Lagrangian
method for the study of test modes in turbulent plasmas. It is similar to
the approach initiated by Dupree \cite{D66}, \cite{D72}. The assumption of
random trajectories with Gaussian distribution limited the application of
Dupree's method to the quasilinear regime. The novelty of our approach
consists of the statistical description of the trajectories, which includes
both random and coherent aspects. The latter lead to non-Gaussian
distribution of trajectories and to complex Lagrangian correlations.

This section is a short review of previous work \cite{VS2013}, \cite{VS2015}%
, \cite{Vlad2013} adapted to the needs of the ISC method.

\bigskip

\subsection{Test modes}

We have studied in \cite{Vlad2013} drift type test modes on turbulent
plasmas. The frequencies and the growth rates are obtained as functions of
the characteristics of the turbulence. They show that ion diffusion
(generated by the random component of the trajectories) has a stabilizing
effect, while ion trapping (the quasi-coherent component) leads to strong
nonlinear effects: increase of the correlation lengths, nonlinear damping of
the drift modes and generation of zonal flow modes. The strength of each of
these processes depends on the parameters of the turbulence.

We use the basic description of the (universal) drift turbulence provided by
the drift kinetic equation in the collisionless and low density limit. We
consider a plasma confined by an uniform magnetic field $B$, taken along the 
$\mathbf{e}_{3}$ axis in a rectagular system of coordinates. The density
gradient with characteristic length $L_{n}$ is taken along $\mathbf{e}_{1}$.
The drift wave instability that is produced by the electron kinetic effects
and the ion polarization drift velocity is studied. The solution of the
dispersion relation for quiescent plasma is (see \cite{GR})%
\begin{equation}
\omega =\frac{k_{2}V_{\ast }}{1+k^{2}\rho _{s}^{2}},\ \ \gamma =\gamma
_{0}\omega \left( k_{2}V_{\ast }-\omega \right)   \label{omfiz}
\end{equation}%
where $\omega $ is the frequency of the mode with wave number $\mathbf{k}%
=[k_{1},\ k_{2},\ k_{z}],~k=\sqrt{k_{1}^{2}+k_{2}^{2}},$ $\mathbf{V}_{\ast
}=V_{\ast }\mathbf{e}_{2},$ $V_{\ast }=T_{e}/(eBL_{n})=\rho _{s}c_{s}/L_{n}$
is the diamagnetic velocity, $\rho _{s}=c_{s}/\Omega _{i},$ $c_{s}=\sqrt{%
T_{e}/mi},$ $T_{e}$ is the electron temperature, $m_{i}$ is the ion mass, $e$
is the absolute value of electron charge and $\Omega _{i}=eB/m_{i}$ is the
ion cyclotron frequency. The constant $\gamma _{0}$ that plays the role of
drive of the instability is $\gamma _{0}=\sqrt{\pi /2}/\left( \left\vert
k_{z}\right\vert v_{Te}\right) ,$\ where $v_{Te}=\sqrt{T_{e}/m_{e}}.$\ 

Electron and ion responses to a small perturbation $\delta \phi $ applied on
the turbulent state with potential $\phi $\ are determined. They lead to a
modified propagator 
\begin{equation}
\overline{\Pi }^{i}=i\int_{-\infty }^{t}d\tau \ M(\tau )\ \exp \left[
-i\omega \left( \tau -t\right) \right] ,  \label{piim}
\end{equation}%
which includes the effects of the turbulence in the function $M(\tau )$
defined by%
\begin{equation}
M(\tau )\equiv \left\langle \exp \left[ i\mathbf{k\cdot }\left( \mathbf{x}%
(\tau )-\mathbf{x}\right) -\int_{\tau }^{t}d\tau ^{\prime }\ \mathbf{\nabla
\cdot u}_{p}\left( \mathbf{x}(\tau ^{\prime })\right) \right] \right\rangle ,
\label{med}
\end{equation}%
where the average $\left\langle {}\right\rangle $ is over the ion
trajectories in the stochastic potential $\phi .$ The polarization drift $%
\mathbf{u}_{p}$ determines a compressibility term due to its divergence $%
\mathbf{\nabla \cdot u}_{p}=-\partial _{t}\Delta \phi /\left( \Omega
_{i}B\right) .$\ The dispersion relation (quasineutrality condition) is the
same as in the quiescent plasma in terms of the propagator%
\begin{equation}
-\left( k_{y}V_{\ast e}-\omega \rho _{s}^{2}k_{\perp }^{2}\right) \overline{%
\Pi }^{i}=1+i\sqrt{\frac{\pi }{2}}\frac{\omega -k_{y}V_{\ast e}}{\left\vert
k_{z}\right\vert v_{Te}}  \label{dr}
\end{equation}%
(see \cite{Vlad2013} for details).

The spectrum of the potential $\phi $ is modeled in accord with the
characteristics of the drift type turbulence. It is zero for $k_{2}=0$
because the drift modes are stable, and it has two symmetrical peaks for $%
k_{2}=\pm k_{0}$\ \ \ \ \ \ 
\begin{equation}
S(\mathbf{k})\sim k_{2}^{m}\exp \left( -\frac{k_{1}^{2}}{2}\lambda
_{1}^{2}\right) \left[ \exp \left( -\frac{(k_{0}-k_{2})^{2}}{2}\lambda
_{2}^{2}\right) -\exp \left( -\frac{(k_{0}+k_{2})^{2}}{2}\lambda
_{2}^{2}\right) \right] .  \label{spectr}
\end{equation}%
The Fourier transform gives the EC

\begin{equation}
E\left( \mathbf{x},t\right) =\Phi ^{2}\partial _{y}^{m}\left[ \exp \left( -%
\frac{x_{1}^{2}}{2\lambda _{1}^{2}}-\frac{x_{2}^{\prime 2}}{2\lambda _{1}^{2}%
}\right) \frac{\sin k_{0}x_{2}^{\prime }}{k_{0}}\right] ,  \label{rEC}
\end{equation}%
where $x_{2}^{\prime }=x_{2}-V_{\ast e}t$ accounts for the motion of the
drift turbulence potential with the diamagnetic velocity. \ The spectrum and
the EC are functions of five parameters: $\Phi ,$ $\lambda _{1},$\ $\lambda
_{2},$\ $k_{0}$ and the power $m$\ that determines the decay of the spectrum
to $k_{2}=0.$\ 

The ISC method shows that the evolution of the drift turbulence is in good
agreement with the model (\ref{spectr}) of the spectrum (see Section 8).

The statistics of the ion trajectories in turbulence with the spectrum of
the type (\ref{spectr}) is discusses in the next subsection. The
micro-probability of the displacements can be Gaussian or non-Gaussian,
depending on the spectrum parameters and on the decorrelation time $\tau
_{d}.$\ More precisely, the statistics depends on the fraction of trapped
ions $n_{tr}$ and on the sizes $s_{1},$ $s_{2}$\ of the structure of
trajectory that appear due to trapping. The most complex case corresponds to 
$n_{tr}$ comparable to the fraction of free trajectories $n_{f}=1-n_{tr}.$
It is characterized by the existence of ion flows determined by ion trapping
in the moving potential. The micro-probability splits in two parts that\ \
move in opposite directions. The trapped trajectories drift with the
potential and have the velocity $V_{\ast }$ and the free trajectory have the
velocity $V_{f}=-nV_{\ast }$\ that ensures zero ion flux (here $%
n=n_{tr}/n_{f})$.\ 

The simple approximation (\ref{probg})\ (discussed in the next Section)
includes the main features of the micro-probability. It has the advantage of
enabling the derivation of simple analytical expressions for the function $M$
(\ref{med}), the renormalized propagator (\ref{piim}) and eventually to
solve the dispersion relation (\ref{dr}) for the complex frequency of the
drift modes (see \cite{Vlad2013}).

The drift modes have modified frequency and growth rate compared to the
quiescent plasmas (\ref{omfiz}), which depend on the characteristics of the
trajectory structures and on the diffusion coefficients%
\begin{equation}
\overline{\omega }_{d}=\frac{1}{2}\left[ \overline{\omega }_{0}+\left(
1-n\right) \overline{k}_{2}+sg\sqrt{\overline{\omega }_{n}^{2}+\frac{4n%
\overline{k}_{2}^{2}}{1+\mathcal{F}\overline{k}^{2}}}\right] ,
\end{equation}%
\begin{equation}
\overline{\gamma }_{d}=\frac{\overline{\gamma }_{0}\left( \overline{\omega }%
_{d}+n\overline{k}_{y}\right) \left[ \left( 1-n\right) \overline{k}_{y}-%
\overline{\omega }_{d}\right] -n_{f}\overline{k}_{i}^{2}\overline{D}_{i}}{%
\left[ \left( 1-n\right) \overline{k}_{y}-\overline{\omega }_{d}\right]
^{2}\left( 1+\mathcal{F}\overline{k}_{\perp }^{2}\right) +n\overline{k}%
_{y}^{2}}\left( \overline{k}_{y}-\overline{\omega }_{d}\right) ^{2},
\label{gamnl}
\end{equation}%
where $\overline{\omega }_{n}=\overline{\omega }_{0}+\left( n-1\right) 
\overline{k}_{2},$ $\overline{\omega }_{0}=\mathcal{F}\overline{k}%
_{2}/\left( 1+\mathcal{F}\overline{k}^{2}\right) $ is the frequency obtained
for $n\rightarrow 0,$ $\mathcal{F=}\exp \left( -s_{i}^{2}k_{i}^{2}/2\right) $%
\ is a factor that depends on the sizes of the trajectory structures, $%
sg=sign\left( \overline{\omega }_{n}\right) $ and $\overline{\gamma }%
_{0}=\gamma _{0}c_{s}/L_{n}.$\ Usual normalized quantities $\overline{k}%
_{i}=k_{i}\rho _{s},$ $\overline{\omega }=\omega L_{n}/c_{s},$ $\overline{%
\gamma }=\gamma L_{n}/c_{s}$\ are used.

The compressibility term determines unstable modes completely different from
the drift modes. They have $k_{y}=0$ and very small frequencies

\begin{equation}
\overline{\omega }_{zf}=-\overline{k}_{x}\overline{a}\frac{1+\mathcal{F}%
\overline{k}_{x}^{2}n_{f}}{1+\mathcal{F}\overline{k}_{x}^{2}},  \label{omzf}
\end{equation}%
\begin{equation}
\overline{\gamma }_{zf}=\frac{n_{tr}\mathcal{F}\overline{k}_{x}^{2}\left[ 
\overline{\gamma }_{0}\overline{\omega }_{zf}^{2}-n_{f}\mathcal{F}\overline{k%
}_{x}^{4}\overline{D}_{x}\right] }{\left( 1+n_{f}\mathcal{F}\overline{k}%
_{x}^{2}\right) \left( 1+\mathcal{F}\overline{k}_{x}^{2}\right) },
\label{gamzf}
\end{equation}%
where $\overline{a}=a/V_{\ast }.$ These unstable modes that are named zonal
flow modes are the consequence of trapping combined with the polarization
drift. The compressibility term in the function $M$ determines correlations
with the displacements 
\begin{equation}
L_{i}(\tau )=-\frac{1}{\Omega _{i}B}\int_{\tau }^{t}d\tau ^{\prime
}\left\langle x_{i}\left( t\right) ~\partial _{\tau ^{\prime }}\Delta \phi
\left( \mathbf{x}(\tau ^{\prime })\right) \right\rangle ,  \label{L}
\end{equation}%
which, as evaluated in \cite{Vlad2013}, can be significant for $i=1$ (in the
direction of the density gradient) for the trapped ions. It can be
approximated by%
\begin{equation}
L_{1}(\tau ,t)\cong a\left( t-\tau \right) ,\ \ a=2\partial _{y}^{2}\Delta E(%
\mathbf{0})\frac{\tau _{fl}V_{\ast }}{\Omega _{i}B^{2}}.  \label{Lxtr}
\end{equation}%
This correlation determines an average velocity for the trapped ions.\ One
can see that when trapping is negligible ($n_{tr}\cong 0),$ $\omega
_{zf}=-k_{x}a$ and $\gamma _{zf}=0,$ and that $\omega _{zf},\gamma _{zf}=0$
for $\overline{a}=0.$

\bigskip

\subsection{Test particles}

The EC of the drift turbulence (\ref{rEC}) has a special shape determined by
the absence of the modes with $k_{2}=0$ in the spectrum (\ref{spectr}).\ It
has negative parts and possibly an oscillating decay as shown in Figure 8b.
This leads to a transport process with important differences compared to the
examples presented in Section 3.2 that are obtained for a monotonically
decreasing EC. The transport coefficients in this type of EC were studied in 
\cite{VS2013}, \cite{VS2015} for ion and electron diffusion. 

We determine here the parameters required by the test mode study and present
a short review of the diffusion process.

The equation for the trajectories in the drift turbulence with a zonal flow
mode potential $\phi _{zf}$ (in the frame that moves with the potential) is

\begin{equation}
\frac{dx_{i}}{dt}=-K_{\ast }^{^{\prime }}\varepsilon _{ij}\partial _{j}\left[
\phi (\mathbf{x},t)+Z~\phi _{zf}(\mathbf{x},t)\right] +V_{d},  \label{eq2}
\end{equation}%
where dimensionless quantities are used with the units: $\rho _{s}$ (for the
distances and the correlation lengths $\lambda _{i},$ $\lambda _{zf}),$\ $%
\tau _{0}=L_{n}/c_{s}$\ (for time) and $\Phi $ (for the potential). In this
normalization, the time flight is a decreasing function of $\Phi $, while $%
\tau _{\ast }$ and $\tau _{d}$ are fixed (independent of the turbulence
amplitude). The normalized amplitude of the zonal flows is $Z=\Phi
_{zf}/\Phi $ and the normalized diamagnetic velocity is\ $V_{d}=1.$ The
dimensionless parameter $K_{\ast }^{^{\prime }}$ is the measure of
turbulence amplitude%
\begin{equation*}
K_{\ast }^{^{\prime }}\equiv \frac{\Phi }{B\rho _{s}V_{\ast }}=\frac{e\Phi }{%
T_{e}}\frac{L_{n}}{\rho _{s}}.
\end{equation*}%
This parameter can be written as $K_{\ast }^{^{\prime }}=K_{\ast }L_{n}/\rho
_{s},$ which shows that it is proportional to the Kubo number that describes
the decorrelation by an average velocity $K_{\ast }=V_{2}/V_{\ast }.$ As
shown in \cite{VS2013}, the transport regimes depend only on $K_{\ast },$
which includes both the effects of the turbulence and of the average
velocity. 

%%%%%%%%%

\begin{figure}[tbp]
\centerline{\includegraphics[height=6cm]{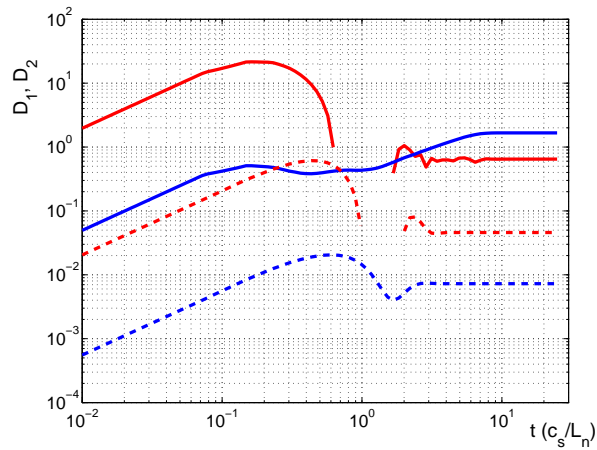}}
\caption{Exemples of diffusion coefficients in drift turbulence: $D_{11}(t)$
(red) and $D_{22}(t)$ (blue) in the quasilinear (dashed lines for $K_{\ast
}^{^{\prime }}=0.5$) and nonlinear (continuous lines for $K_{\ast
}^{^{\prime }}=5$) regimes. The other parameters are $k_1^0=2.5,$ $\protect%
\lambda_1=2,$ $\protect\lambda_2=1,$ $V_d=1$ and $\protect\tau_d=5.$ }
\label{Figure3}
\end{figure}

%%%%%%%%%

Typical results for the diffusion coefficients are shown in Figure 3. We
note that the transport is not isotropic. The amplitudes of the velocity
components are different for the EC (\ref{rEC}). The special shape of the EC
leads to diffusion regimes for $D_{1}$ that have similar dependence on the
decorrelation time in the quasilinear and nonlinear regime. This determines
smaller values of the quasilinear $D_{1}$ than in a decaying EC.\ The shape
of the EC also determines negative parts in the time dependence of $D_{11}(t)
$ (seen as breaks in the red lines in Figure 3). The modification of $D_{2}$%
\ appears in the quasilinear regime too, and consists of significant
decrease, as seen by comparing the dashed blue curves in Figures 3 and 2.

A large increase of the asymptotic coefficient $D_{2}$ appears at the
transition from the quasilinear to the nonlinear regime, which provides a
strong auto-control parameter. As seen in Figure 3, the ratio of the
asymptotic values in the nonlinear and quasilinear regime is much larger for 
$D_{2}$ than for $D_{1}.$ \ 

%%%%%%%%%

\begin{figure}[tbp]
\centerline{\includegraphics[height=6cm]{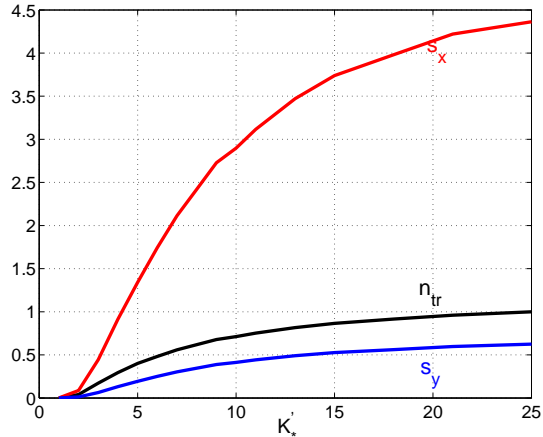}}
\caption{The characteristics of the trajectory structures as functions of
the normalized amplitude of the potential $K_{\ast}^{^{\prime }}$ : the
fraction of trapped ions (black) and the size of the structures $s_x$ (red), 
$s_y$ (blue). The other parameters are as in Figure 3. }
\label{Figure4}
\end{figure}

%%%%%%%%%

\bigskip

The parameters of the trajectory structures are obtained using the DTM as
weighted averages of the decorrelation trajectories $\mathbf{X}^{S}(t)$ in
the static potential. The solutions of Eq. (\ref{DTprim}) are periodic
functions in this case with the periods $T(\phi ^{0},\mathbf{v}^{0})$ that
depend on the subensemble. The fraction of trapped trajectories at time $t$
is%
\begin{equation}
n_{tr}(t)=\int d\phi ^{0}P(\phi ^{0})\int
dv_{1}^{0}dv_{2}^{0}P(v_{1}^{0})P(v_{1}^{0})\ c_{tr}(t;\phi ^{0},\mathbf{v}%
^{0}),  \label{ntr}
\end{equation}%
where $c_{tr}(t;\phi ^{0},\mathbf{v}^{0})=1$ if $t>T(\phi ^{0},\mathbf{v}%
^{0})$\ and $c_{tr}(t;\phi ^{0},\mathbf{v}^{0})=0$ if $t<T(\phi ^{0},\mathbf{%
v}^{0}).$\ The sizes of the trajectory structures in the two directions are 
\begin{equation}
s_{i}(t)=\int d\phi ^{0}P(\phi ^{0})\int
dv_{1}^{0}dv_{2}^{0}P(v_{1}^{0})P(v_{1}^{0})\ X_{i}^{\max }(t;\phi ^{0},%
\mathbf{v}^{0}),  \label{Si}
\end{equation}%
where $X_{i}^{\max }(t;\phi ^{0},\mathbf{v}^{0})$ is the dimension of  $%
\mathbf{X}^{S}(t;\phi ^{0},\mathbf{v}^{0})$ along the direction $i$ if $%
t>T(\phi ^{0},\mathbf{v}^{0})$ and $X_{i}^{\max }(t;\phi ^{0},\mathbf{v}%
^{0})=0$ if $t<T(\phi ^{0},\mathbf{v}^{0}).$

The functions $n_{tr}(t)$ and $s_{i}(t)$\ describe the growth of the
trajectory structures. These functions saturate in a time $\tau _{s},$ which
is the characteristic time for the formation of the structure. $\tau _{s}$
is an increasing function of the amplitude of the turbulence $K_{\ast
}^{^{\prime }}.$ The asymptotic values of $n_{tr}(t)$ and $s_{i}(t)$\
describe the parameters of the structures as functions of the
characteristics of the turbulence. The dependence on the amplitude of the
turbulence is presented in Figure 4. It shows that the structures
continuously increase when turbulence amplitude increases and eventually\
they include all the trajectories $n_{tr}\rightarrow 1$. In the presence of
a decorrelation process with characteristic time $\tau _{d}<\tau _{s}$, the
parameters of the structures are smaller because they are destroyed during
the formation at the time $\tau _{d}.$ They can be approximated by $%
n_{tr}(\tau _{d})$ and $s_{i}(\tau _{d}),$\ instead of the asymptotic
values.\ 

\bigskip

The micro-probability $P^{\tau _{d}}(\mathbf{x},t)$ can be evaluated using
the DTM as a histogram of the decorrelation trajectories. It can be
Gaussian, or non-Gaussian, depending on turbulence parameters and on the
decorrelation time $\tau _{d}.$\ Non-Gaussian $P^{\tau _{d}}$\ are
associated to the presence of trapping. The micro-confinement appears very
clearly in the probability as a narrow peak in $\mathbf{x=0},$\ which
remains invariant after a formation time $\tau _{s},$\ the same as in the
time evolution of the fraction of trapped trajectories $n_{tr}(t).$ $P^{\tau
_{d}}$\ has also a time dependent part around this maximum, which is
determined by the free particles. For the drift turbulence case, the motion
of the potential with the diamagnetic velocity leads to the separation of
the two parts of the micro-probability: the narrow peak is displaced in the
direction $V_{d}\mathbf{e}_{2}$ and the free particle part concentrates in
the opposite direction then moves. This result is in qualitative agreement
with the process of flow generation in zero divergence velocity fields
(Section 2), but the flow velocities obtained with the DTM are smaller than (%
\ref{flow}).

The analytical results for the test modes (Section 4.1) are obtained using
simplified approximations of $P^{\tau _{d}}$\ in agreement with Eq. (\ref%
{flow}) for the ion flows (see \cite{Vlad2013} for details). The trapped
particle part was represented for simplicity by a Gaussian function, but
with fixed dispersion given by the size $s_{i}$ of the structures $%
S_{i}=s_{i}^{2}$. The shape of this function does not change much the
estimations. The free trajectories are described by a Gaussian with
dispersion that grows linearly in time: $S_{i}^{\prime }\left( \tau \right)
=S_{i}+2D_{i}\left( t-\tau \right) ,$ $i=x,y$. The initial value $%
S_{i}^{\prime }\left( t\right) =S_{i}$ is an effect of trapping. It
essentially means that the trajectories are spread over a surface of the
order of the size of the trajectory structures when they are released by a
decorrelation mechanism. Introducing the flows (\ref{flow}), the
approximation for the micro-probability is\ 
\begin{equation}
P^{\tau _{d}}(x,y,t)=n_{tr}G(x,y-V_{d}t;\mathbf{S})+n_{f}G(x,y-V_{f}t;%
\mathbf{S}^{\prime }),  \label{probg}
\end{equation}%
where $G(\mathbf{x};\mathbf{S})$ is the 2-dimensional Gaussian distribution
with dispersion $\mathbf{S}=(S_{x},S_{y}).$ We note that this probability is
non-Gaussian.

The separation of the distribution and the existence of ion flows in drift
type turbulence are confirmed by numerical simulations \cite{Jenko}.

\section{The iterated self-consistent method}

The idea of the iterated self-consistent (ISC) method is based on the
difference between the description of turbulence evolution and the mode
representation. One can define in the evolution equation characteristic
times for each process (represented by a term in the equation). The ordering
of these characteristic times shows which are the processes that have
important contributions at a given stage of evolution. Neglecting the small
terms leads to simplified equations that give the short time evolution, for
time smaller than the characteristic times of the neglected terms. The
estimation of the characteristic times provides the basis for a systematic
approximations that can strongly simplify the evolution equation at small
time. On the other side, the mode representation of turbulence does not deal
with time evolution but it determines the possible frequencies and wave
numbers that are supported by the system and their tendency of amplification
or damping. These quantities include effects of processes that are
negligible at short time.

The combination of the study of test modes on turbulent plasma with the
evaluation of the short time equilibrium of the distribution functions\ can
provide a much simplified approach and even the possibility to develop a
semi-analytical approximate method for the turbulence evolution. \ 

\ More precisely, the short time solution is an approximation that is not
valid at large times where the neglected small terms in the evolution
equation determine the accumulation of significant effects. These effects
are taken into account using the test modes. The frequency and the growth
rate of a small perturbation $\delta \phi $ of the background potential $%
\phi $ include the terms and the corresponding effects that are neglected in
the small time equilibrium distribution. The effective calculation of these
characteristics of the modes requires the statistics of the trajectories
(for evaluating the average propagator). The diffusion coefficients and the
probability of displacements depend on the momentary Eulerian correlation of
the background potential. The latter is provided by the small time solutions
of the evolution equations. Finally, the frequencies and the growth rates of
all modes determine the evolution of the background potential.

Thus the methodology of the ISC approach consists of repeated steps that
contain the following calculations.

\textbullet \qquad The evaluation of approximate (short time) equilibrium
distribution functions in the presence of background turbulence and of the
Eulerian correlation of the potential.

\textbullet \qquad The calculation of the statistical characteristics of the
trajectories (diffusion coefficients, probability of displacements,
characteristics of the quasi-coherent structures) as functions of the
Eulerian correlation of the potential. This is one of the main modules,
presented in Section 4.2.

\textbullet \qquad The calculation of the renormalized propagator (averaged
over trajectories) and evaluation of the frequencies and the growth rates of
the test modes. This is the second main module of the ISC.

\textbullet \qquad The evolution of the spectrum on a small time interval is
obtained using the growth rates of all modes. It is the starting point of a
new step in this iterated method.

\bigskip

In the case of the drift instability studied here \cite{GR}, the small time
distributions are the adiabatic response for the electrons and $%
f_{0}^{i}=n_{0}(x_{1})F_{M}^{i}\left( 1+e\phi (\mathbf{x}-V_{d}t\mathbf{e}%
_{2})/T_{e}\right) $ for the ions, where $n_{0}(x_{1})$\ is plasma density, $%
V_{d}$ is the diamagnetic velocity and $F_{M}^{i}$\ is the Maxwell
distribution of ion velocities. They correspond to the stable drift waves
and show that an arbitrary potential $\phi $ in a nonuniform plasma drifts
with the diamagnetic velocity. The neglected terms are taken into account in
the linearized equation for the small perturbation. The kinetic effects of
the electrons determine their nonadiabatic response, which is the drive of
the instability. This response is the same as in quiescent plasmas due to
the fast parallel motion of the electrons. The compressibility effects
determined by the polarization drift of the ions make the growth rates
positive. The stochastic ion trajectories in the background potential $\phi $%
, which can become rather complex by acquiring coherent aspects at large
amplitudes of $\phi ,$ change the frequencies and the growth rate leading to
strong nonlinear effects in the evolution of turbulence.

Several important simplification can be operated in methodology of the ISC
method. The most important concern the test mode module that is practically
eliminated in the present case by the analytical results available for the
frequencies and the growth rates of drift and zonal flow modes in turbulent
plasmas (Section 4.1). Moreover, we have found that the spectra of the drift
and zonal flow components obtained in the iterated evolution can be
approximated by expressions of the type of Eq. (\ref{spectr}). It is thus
sufficient to determine the parameters in this expression at each step. This
permitted to have analytical expressions for the EC in the test particle
module.

A computer code was developed for the study of the drift turbulence using
the ISC method. We note that this code is not performing the simulation of
the turbulence, but it calculates the quantities related to test particles
and test modes according the iterated procedure presented here. The run time
is of few hours on a laptop.

\section{Drift turbulence evolution}

There are two possible mechanisms for the attenuation of the drift
turbulence.

The first is the diffusive damping, which can be strongly enhanced due to
the increase of $D_{2}$ in the nonlinear regime. The process is produced by
the diamagnetic velocity that modifies the total potential producing bundles
of open contour lines with average orientation along its direction. The
zonal flow modes can contribute to this process since they provide an
additional component along $V_{d}\mathbf{e}_{2}$ \cite{VS2013}. This
velocity is stochastic with zero average,\ but it determines an average
effect due to the nonlinear dependence of $n_{tr}$\ on $V_{d}$\ (Figure 4).

The second mechanism is the non-linear self-damping determined by the ion
flows. The increase of the rate of trajectory trapping $n=n_{tr}/(1-n_{tr})$
determine the decrease of $\gamma _{d},$\ which becomes negative for all
values of the wave numbers at $n=1,$\ as results from Eq. (\ref{gamnl}).

The self-consistent evolution is expected to be much more complicated since
the processes identified by a separate analysis of the two main modules can
be simultaneous present or they can combine in a synergistic way. Moreover,
there are other parameters that are included in both modules (the
correlation lengths, the dominant wave numbers, the size of the trajectory
structures) that change during the evolution and can strongly affect the
quasi-coherent structures.

We have found two types of evolution depending on the strength of the drive
of the instability $\overline{\gamma }_{0}$. The evolution at small drive
(small $\overline{\gamma }_{0}$ that corresponds to large $k_{z})$\ is
controlled by the diffusion. At high drives that appear for smaller $k_{z}$,
the effects of trajectory trapping become dominant. In both cases, there is
a similar initial stage.

The evolution of drift turbulence is analyzed below. We use in the figures
red and blue colors for the two-dimensional quantities (red for the
direction $\mathbf{e}_{1}$ of the density gradient and blue for the
direction $\mathbf{e}_{2})$ and black for the scalar functions or for the
zonal flows.\ \ \ 

\subsection{The initial stage}

The evolution of turbulence has an initial stage that depends on the initial
condition for the spectrum. However, for all physically reasonable
conditions, the shape of the spectrum changes and it always develops two
symmetrical maxima in $k_{1}=0$ and $k_{2}=\pm k_{0}.$\ They are produced by
the diffusive damping\ of the modes, which acts very strongly on large $%
k_{2} $\ and $k_{1}\cong 0$ domain and makes $\gamma <0$\ even at small
values of the diffusion coefficient $D_{y}.$ The diffusive decay at large $%
\left\vert k_{2}\right\vert $ combined with the growth rate that decay to
zero at small $\left\vert k_{2}\right\vert $ determine the generation of the
two peaks in the spectrum.\ This leads to the increase of the anisotropy of
the turbulence. The amplitude of the normalized velocity $V_{1}$ increases
while $V_{2}$ decreases in this stage (see the small time evolution in the
Figures 5a and 6a). The cause of this behaviour is the increase of the
dominant wave number $k_{2}^{0}$ and of the correlation lengths $\lambda
_{i} $ (Figures 5b and 6b). The amplitude of the potential fluctuations
represented by $K_{\ast }^{^{\prime }}$ increases exponentially.

It is interesting to note that, besides the large stochastic velocity along
the density gradient ($V_{1}\gg V_{2}),$ the diffusion coefficient is
smaller ($D_{1}>D_{2}).$ This behavior is the effect of the special shape of
the EC of the drift turbulence that yields decreasing diffusion coefficient $%
D_{1}$ at large $\tau _{d}$\ even in the quasilinear regime.\ \ 

The initial stage is not a regime of completely independent evolution of the
modes. The influence of the background turbulence that consists of the
diffusive damping is not negligible at the very small values that correspond
to this stage. Namely, the growth rate in the quasilinear limit that can be
obtained from Eq. (\ref{gamnl}) for $n=0$\ and $\mathcal{F}=1$ at $k_{1}=0$
is%
\begin{equation*}
\overline{\gamma }_{d}=\overline{k}_{2}^{2}\left( \gamma _{0}\frac{\overline{%
k}^{2}}{(1+\overline{k}^{2})^{2}}-\overline{D}_{2}\right) .
\end{equation*}%
The parentheses becomes negative at large enough $\overline{k}$\ because the
first term decays to zero.

\subsection{Evolution at weak drive}

Typical evolution at weak drive is presented in Figure 5, where $\overline{%
\gamma }_{0}=1.$\ After the initial stage, the diffusive damping becomes
stronger and it determines important changes in the evolution. The amplitude
of the potential $K_{\ast }^{^{\prime }}$\ continuous to increase but with a
slower rate: the exponential evolution is replaced by a roughly linear
increase (Figure 5a, the back curve for $K_{\ast }^{^{\prime }}$). Both
components of the normalized velocity decrease. They have a fast decay
followed by the tendency of saturation (Figure 5a). This is the effect of
diffusional damping. The diffusion coefficients increase as $\left( K_{\ast
}^{^{\prime }}\right) ^{2}$ and they produce the decrease of the growth rate
of the drift modes.

%%%%%%%%%

\begin{figure}[tbp]
\centerline{\includegraphics[height=5cm]{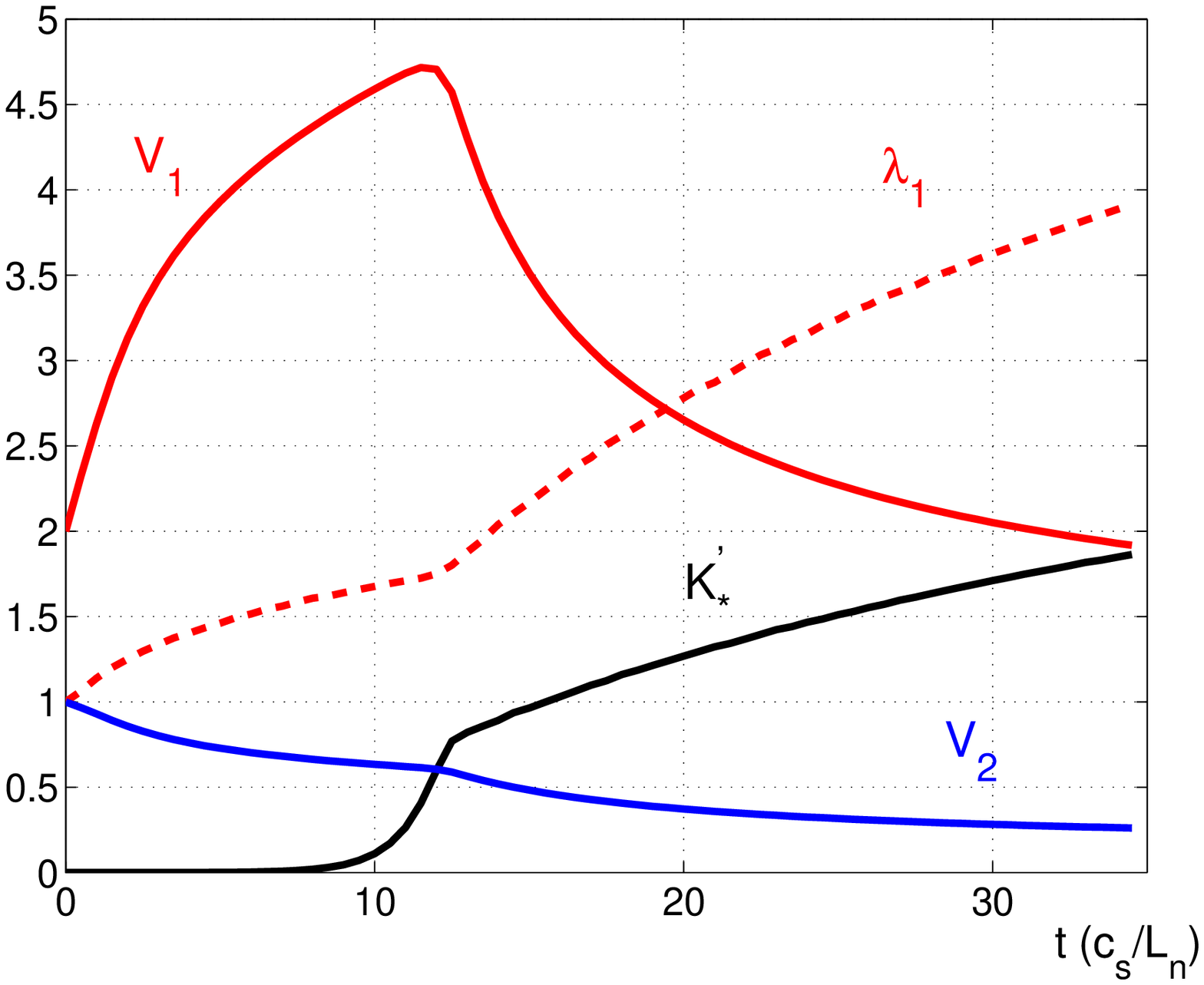}} \centerline{%
\includegraphics[height=5cm]{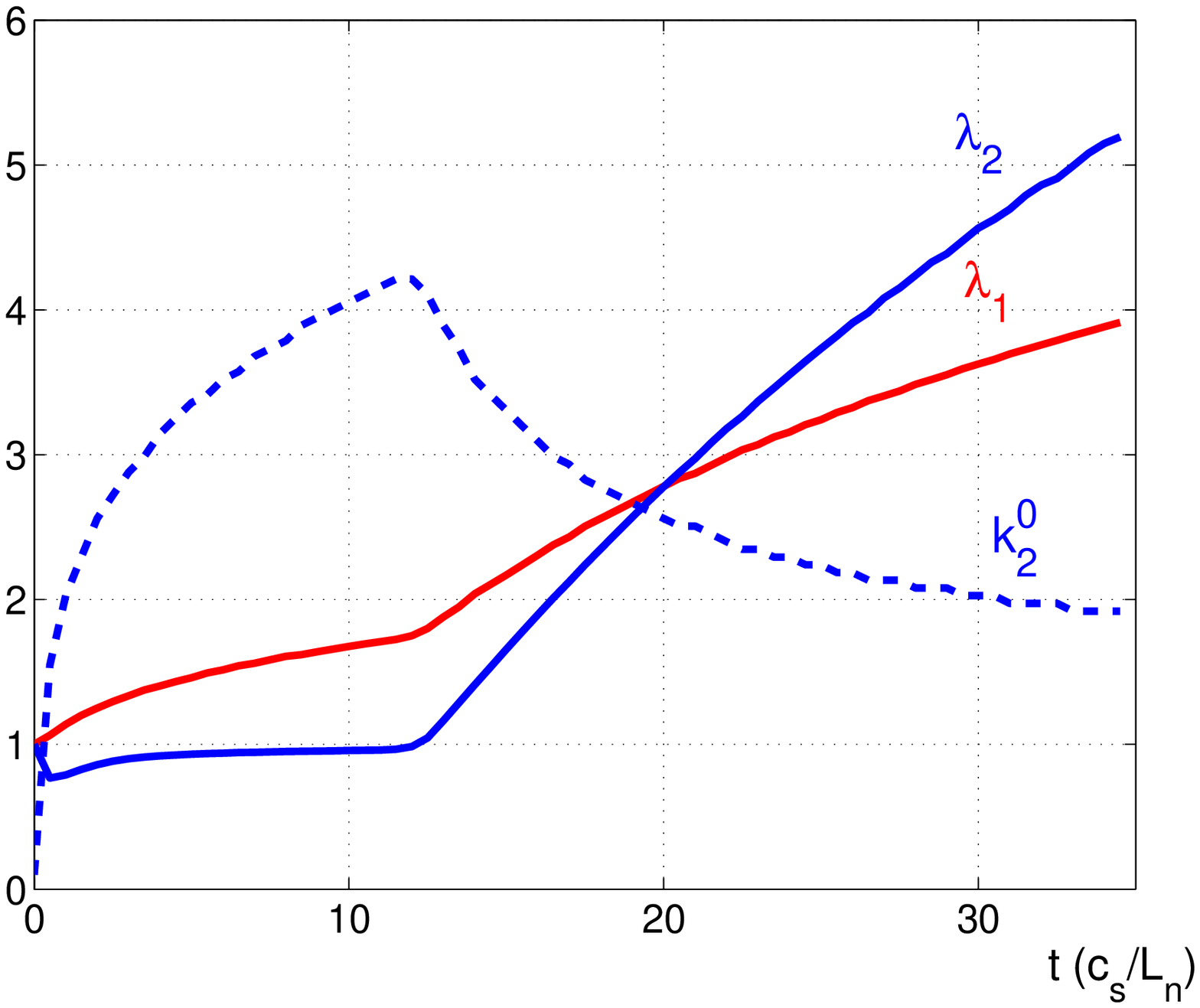}} \centerline{%
\includegraphics[height=5cm]{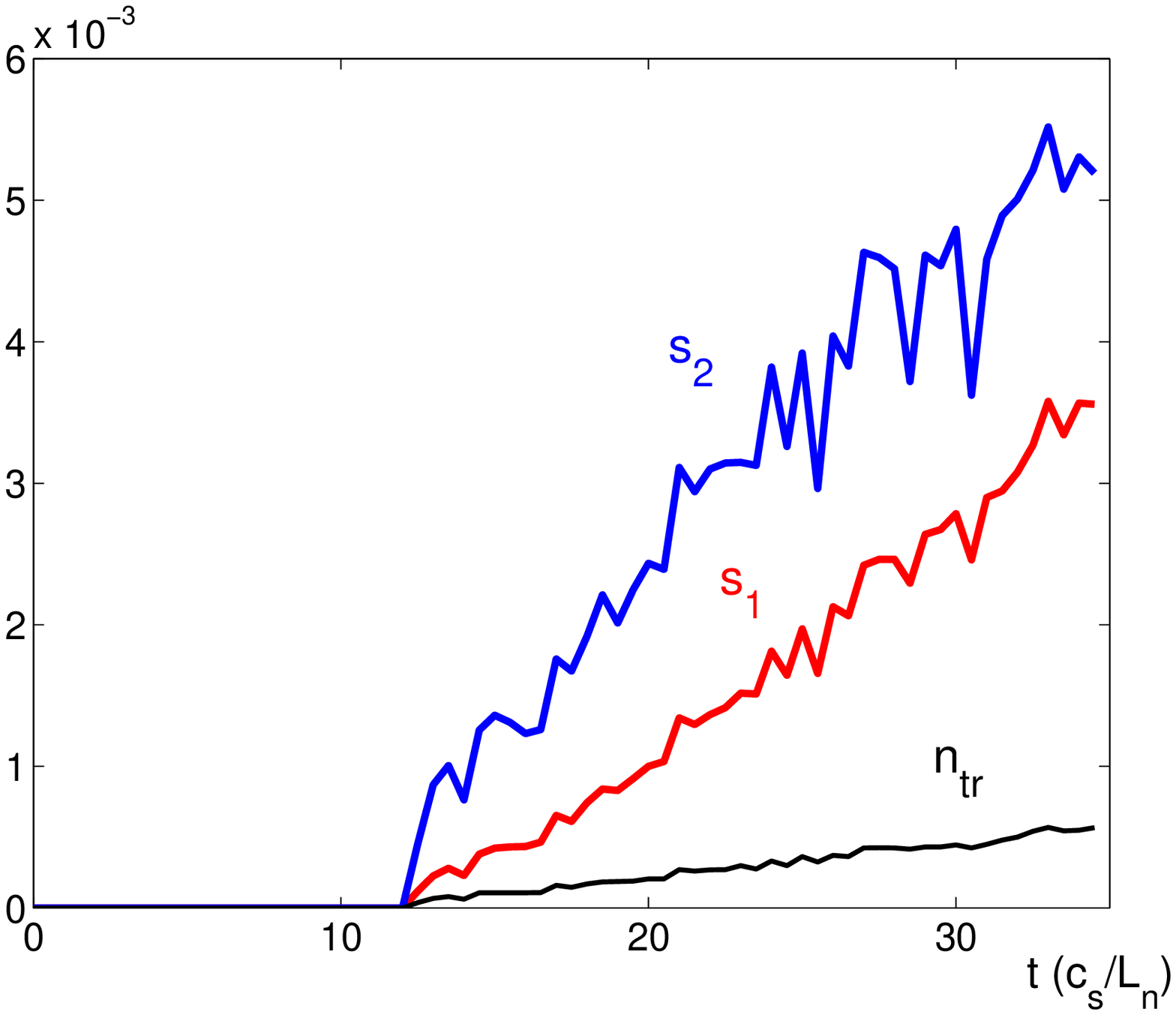}}
\caption{Drift turbulence evolution for weak drive. a) The turbulence
amplitude parameter $K_{\ast }^{^{\prime }}$ (black line) compared to the
nonlinear limit $\protect\lambda_1$. The components of the normalized
velocity are also plotted (red and blue curves). b) The parameters of the
EC. c) The trajectory structure sizes $s_1$, $s_2$ and the fraction of
trapped trajectories $n_{tr}$. }
\label{Figure5}
\end{figure}

%%%%%%%%%

Besides this well known effect of damping, the diffusion determines
significant changes of the shape of the spectrum. The width of the two peaks
decreases on both directions, which determines the increase of the
correlation lengths. It also determines the displacement of the position of
the peaks toward smaller wave numbers. The evolution of the spectrum
parameters is shown in Figure 5b.

The condition for the transition to the trapping regime ($K_{\ast
}^{^{\prime }}>\lambda _{1})$ is not attained in this case since, as seen in
Figure 5a, the increase rate is larger for $\lambda _{1}$ than for $K_{\ast
}^{^{\prime }}.$ The structures of trajectories are practically absent, as
seen in Figure 5c where $n_{tr}<10^{-3}$ and the $s_{i}<10^{-2}.$\ 

Drift turbulence evolution at weak drive is essentially determined by ion
diffusion, which reduces the growth rate and determines a significant change
of the spectrum that prevents the development of the coherent effects
produced by trapping. \ \ \ 

\subsection{Evolution at strong drive}

%%%%%%%%%

\begin{figure}[tbp]
\centerline{\includegraphics[height=5cm]{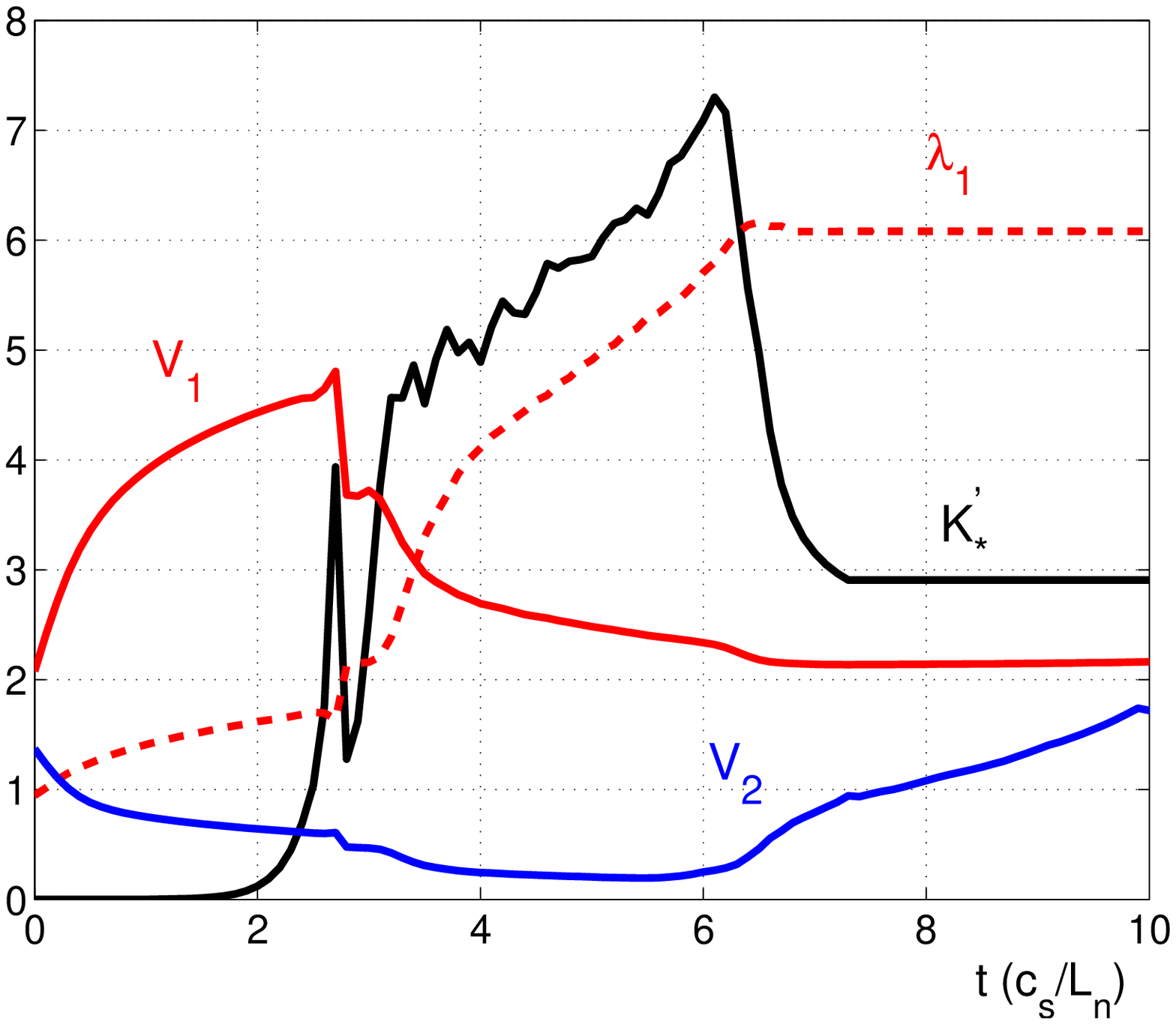}} \centerline{%
\includegraphics[height=5cm]{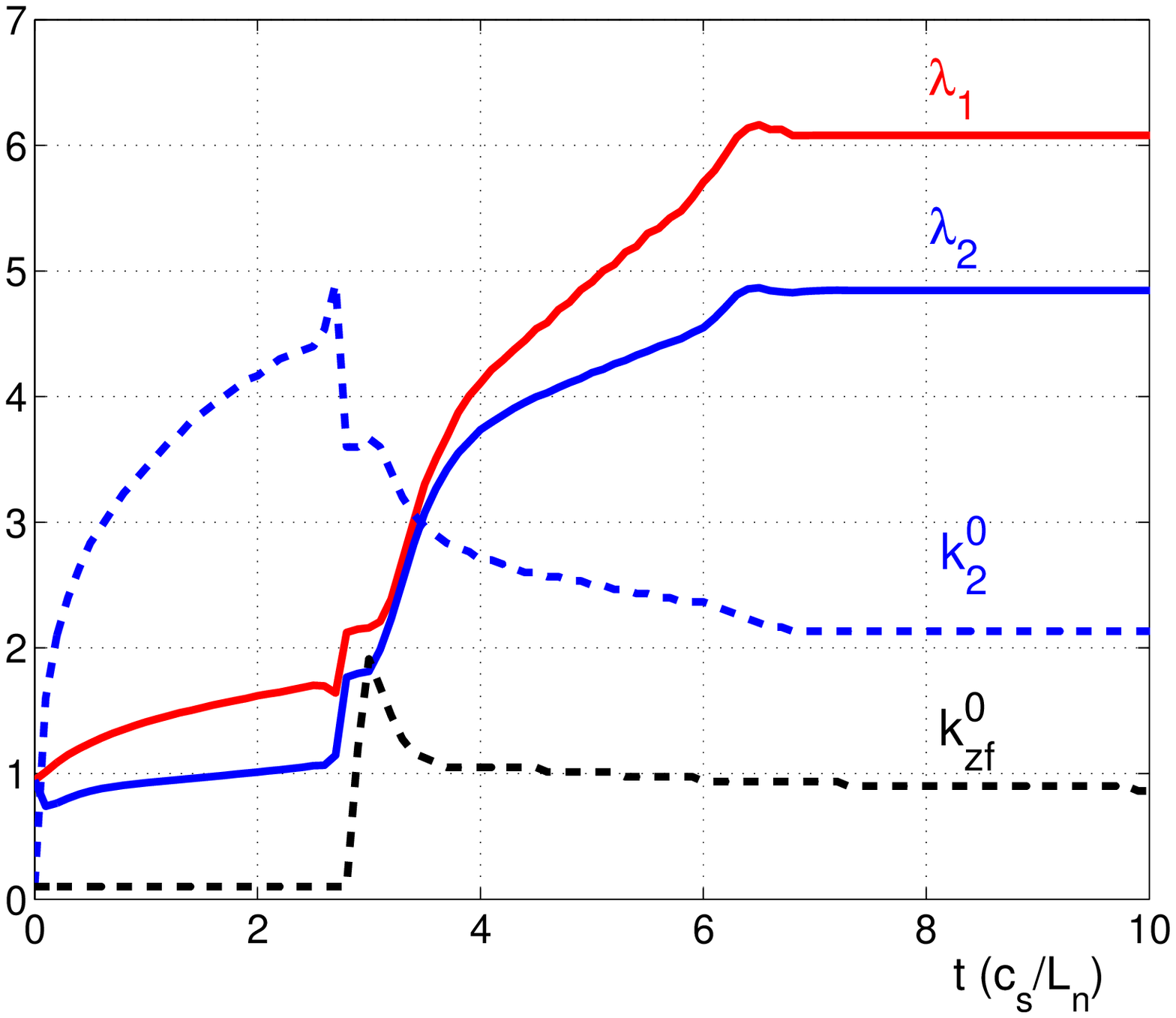}} \centerline{%
\includegraphics[height=5cm]{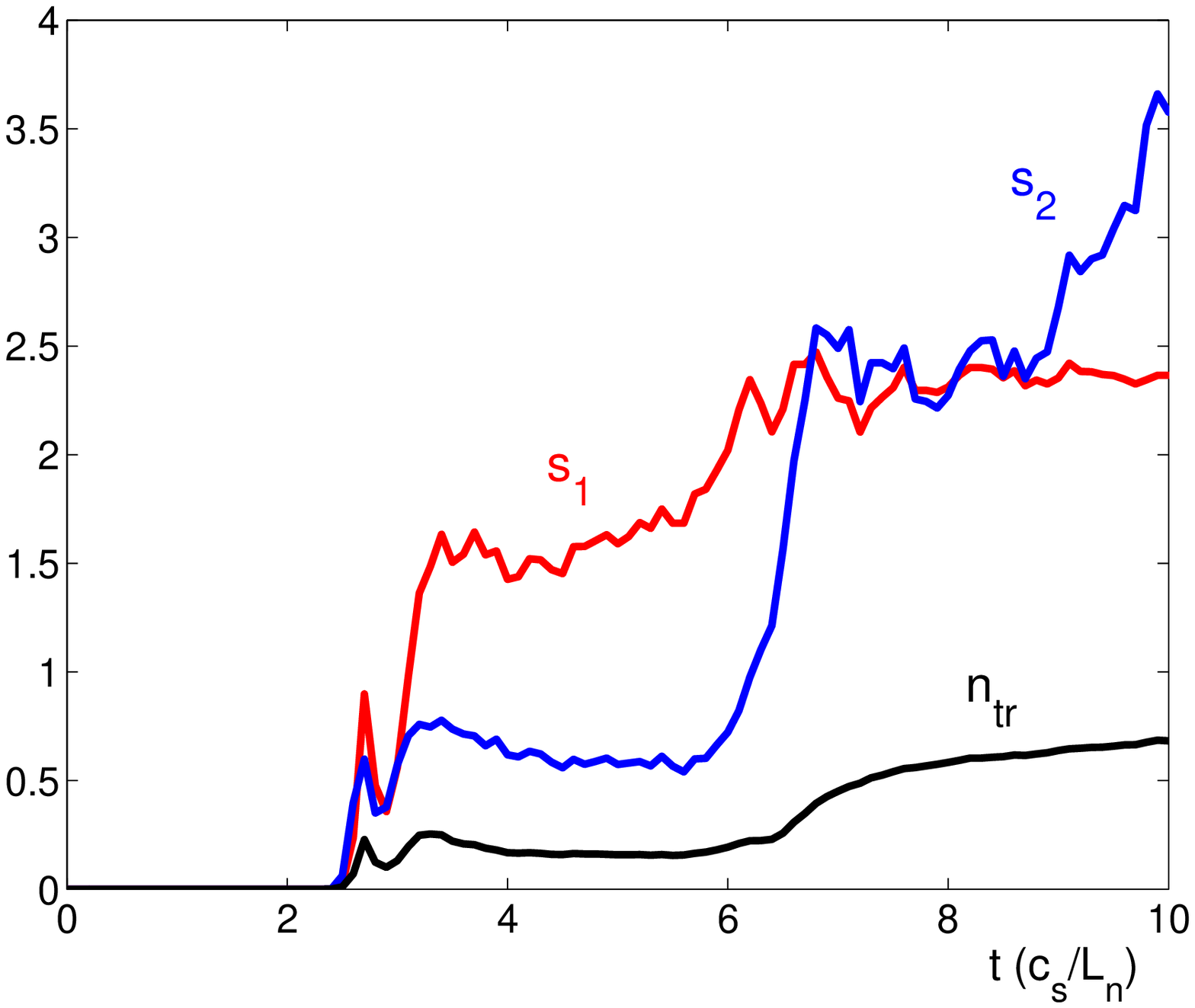}}
\caption{The same as in Figure 5, but for the evolution of drift turbulence
with strong drive. The dominant wave number of the zonal flow modes is shown in b (black dashed line). }
\label{Figure6}
\end{figure}

%%%%%%%%%

%%%%%%%%%

\begin{figure}[tbp]
\centerline{\includegraphics[height=5cm]{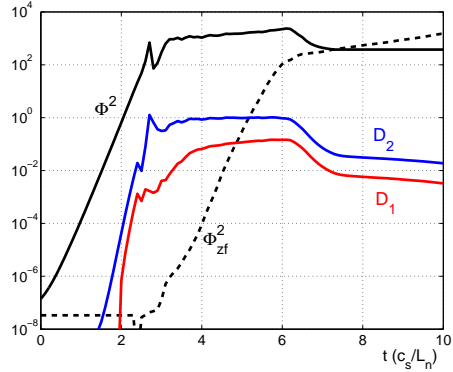}}
\caption{Evolution of the amplitude of the drift and zonal flow modes
compared with the evolution of the diffusion coefficients.}
\label{Figure7}
\end{figure}

%%%%%%%%%

The increase of the drive parameter determines a strong effect on turbulence
evolution, as seen in Figures 6-8, where $\overline{\gamma }_{0}=5.$ This
large values of $\overline{\gamma }_{0}$ lead, since the initial stage of
the evolution, to a faster increase of the amplitude of the normalized
potential $K_{\ast }^{^{\prime }}$. Ion diffusion determines the same
effects as in the weak drive case, but the fast exponential increase of the
potential up to larger values makes $K_{\ast }^{^{\prime }}>\lambda _{1},$
and determines ion trapping.

Three time intervals with different characteristics of the evolution can be
identified in the typical example presented in Figure 6: $t<2.5,$ $2.5<t<6$
and $t>6.$\ The first interval corresponds to the initial stage of
quasi-independent development of the modes. The other two intervals are
influence by the trapping process but in different ways.

As seen in Figure 6a, the evolution enters in the nonlinear regime $K_{\ast
}^{^{\prime }}>\lambda _{1}$ at the end of the initial stage determining ion
trajectory trapping (Figure 6c). The potential $K_{\ast }^{^{\prime }}$ has
a fast transitory decay, followed by a regime of continuous increase of the
average $K_{\ast }^{^{\prime }}$ (Figure 6a, black curve). During this
stage, $K_{\ast }^{^{\prime }}>\lambda _{1}$ and both parameters increase
with roughly the same rate. The ratio $K_{\ast }^{^{\prime }}/\lambda _{1}$
is approximately constant, which corresponds to small variation of the
parameters of the trajectory structures (Figure 6c, the time interval $[4,\
6]$). The spectrum and the EC are strongly modified during this interval.
The correlation lengths $\lambda _{i}$ increase significantly and the
dominant wave number $k_{2}^{0}$\ decreases (Figure 6b).\ The evolution of
the velocity is different from that of the potential as reflected in the
normalized velocities $V_{1},~V_{2},$ which are time dependent functions
(see Figure 6a). Their evolution is actually determined by the change of the
characteristics of the EC, because $V_{1}^{2}\cong \left( k_{2}^{0}\right)
^{2}+3/\lambda _{2}^{2}$ and $V_{2}^{2}\cong 1/\lambda _{1}^{2}.$

A different evolution regime appears at $t>6$\ \ in the example presented in
Figure 6. The normalized potential drops and it saturates (Figure 6a). The
normalized velocity $V_{1}$ saturates and $V_{2}$ increases. The
characteristics of the EC, $\lambda _{i}$\ and $k_{2}^{0},$ also saturate
(Figure 6b). The condition for the existence of trajectory structures is no
more fulfilled ($K_{\ast }^{^{\prime }}$ is significantly smaller than $%
\lambda _{1})$. However, trapping becomes much stronger ($n_{tr}$ and the
size of the structures increase, as seen in Figure 6c).

This rather nontrivial evolution of the drift turbulence can be understood
by analyzing the nonlinear effects that influence the growth rate (\ref%
{gamnl}): the ion diffusion, the formation of ion trajectory structures and
flows. \ \ 

%%%%%%%%%

\begin{figure}[tbp]
\centerline{\includegraphics[height=5cm]{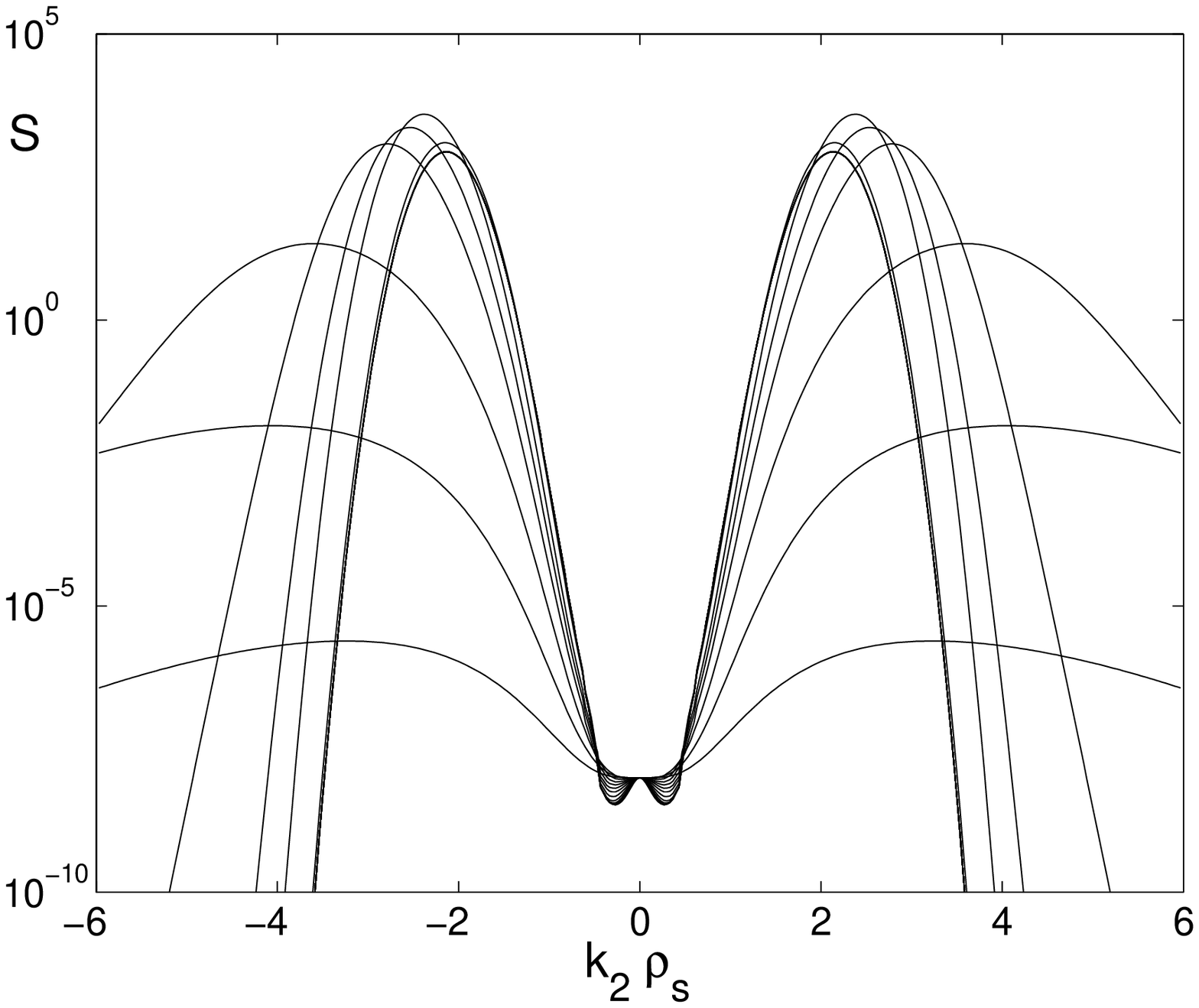}} \centerline{%
\includegraphics[height=5cm]{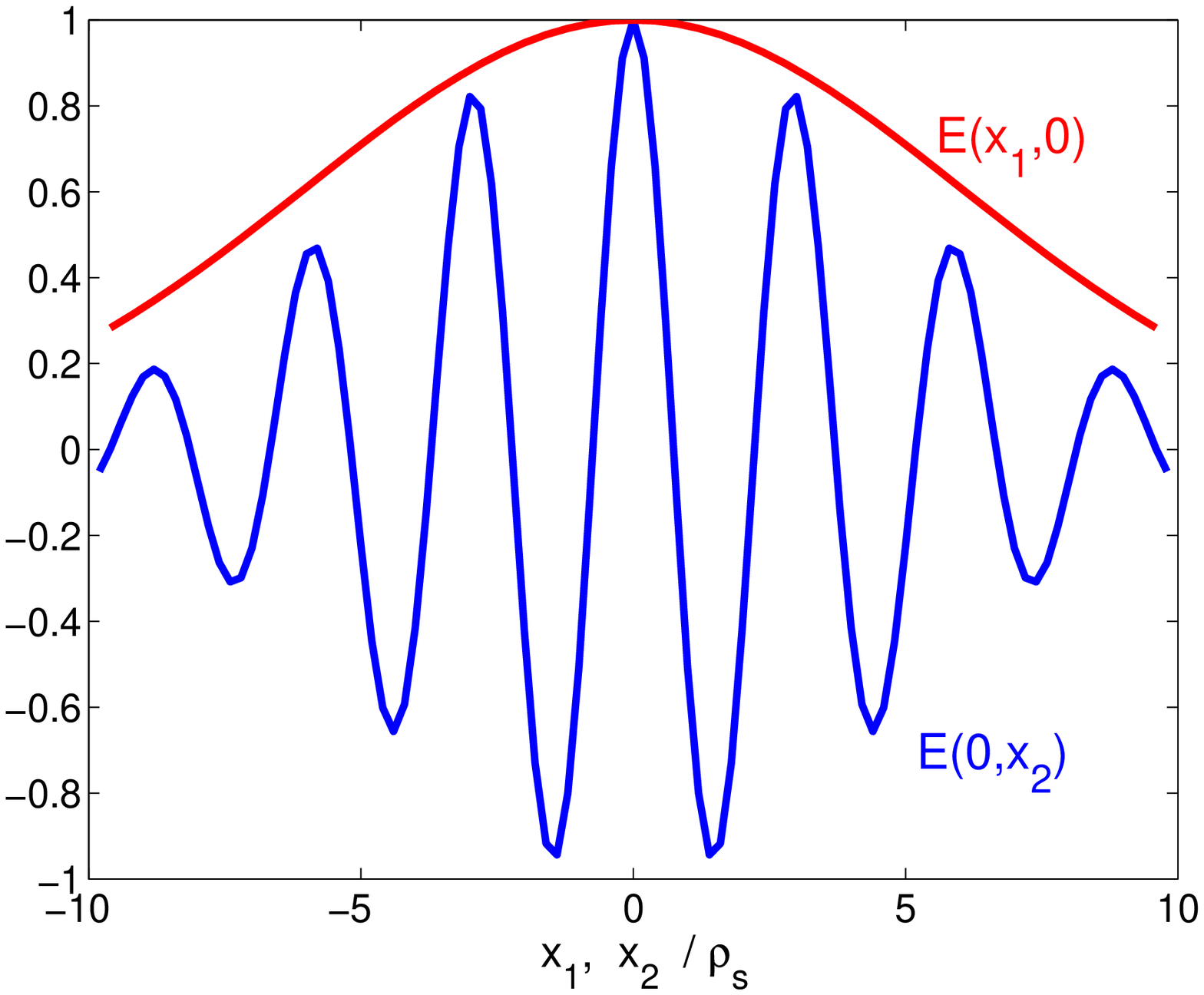}} \centerline{%
\includegraphics[height=5cm]{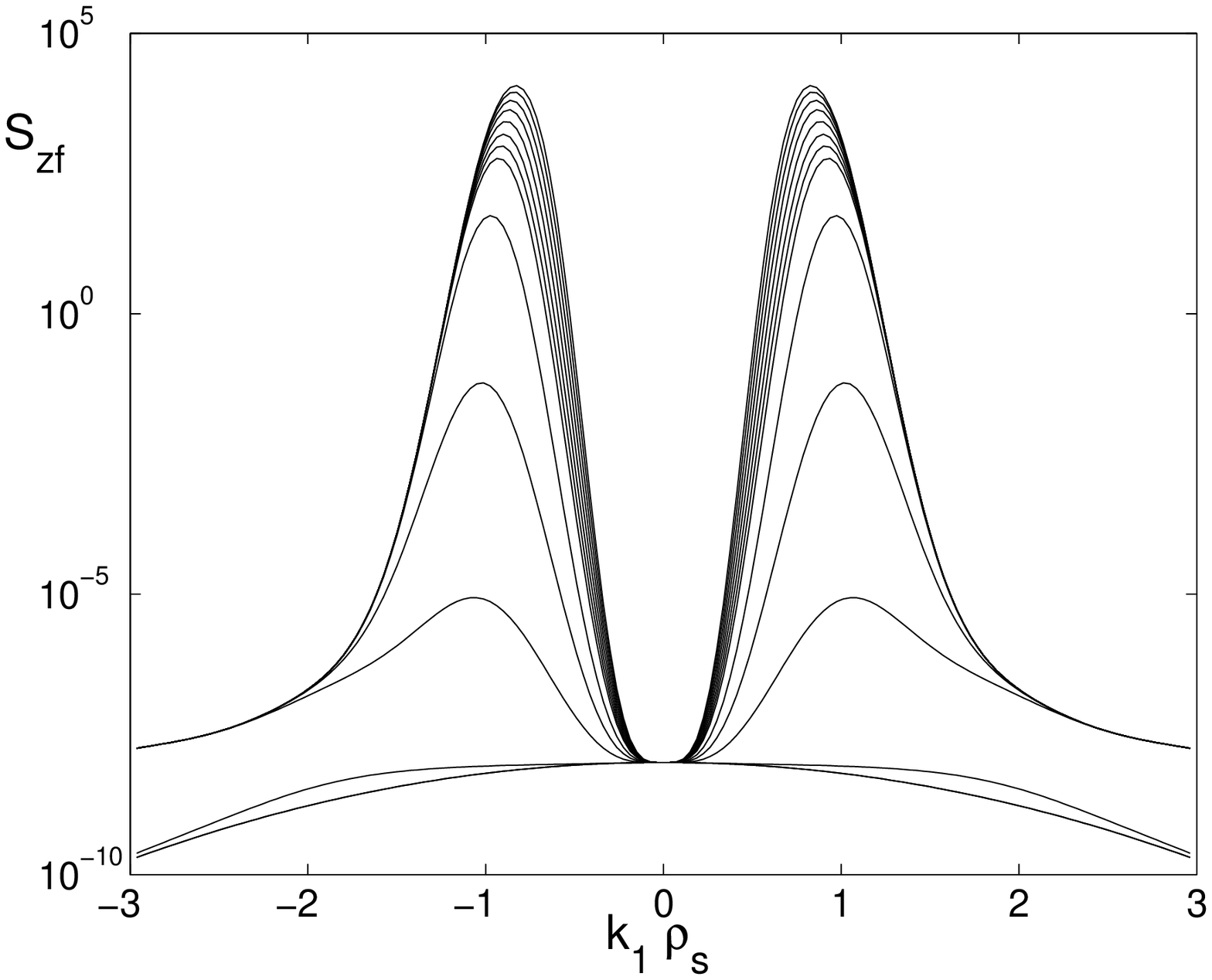}}
\caption{The spectrum (a) and the EC (b) of the drift turbulence and the
spectrum of the zonal flow modes (c). The spectra are plotted at time
intervals $\delta t=L_n/c_s$ and the EC is at $t=9 L_n/c_s$. }
\label{Figure8}
\end{figure}

%%%%%%%%%

The evolution of the diffusion coefficients $D_{i}$ is presented in Figure
7, where the amplitudes of the drift turbulence $\Phi ^{2}$ and of the zonal
flow potential $\Phi _{zf}^{2}$\ are also shown. One can see that the zonal
flow modes become unstable when trapping appears and they grow exponentially
until $t\cong 6.$ Their amplitude is negligible compared to that of the
drift turbulence during this time interval. Thus, it is not expected an
important role of the zonal flow modes at this stage.

The diffusion coefficient $D_{2}$ is much larger than $D_{1}.$ It has a very
large increase rate during the first transition into the nonlinear regime.
The cause is ion trapping in the potential that drifts with the diamagnetic
velocity, which determines a strong amplification of the diffusion along $%
V_{d}\mathbf{e}_{2},$ as discusses in Section 4.2 and represented in Figure
3. The strong diffusion determines the transitory decay of the potential $%
K_{\ast }^{^{\prime }}$ under the nonlinearity limit. Consequently, the
structures of the ion trajectories are destroyed. The fraction of trapped
trajectories and the size of the structures decrease when $K_{\ast
}^{^{\prime }}$ decreases (Figure 4). The diffusion coefficient $D_{2}$
decreases due to both effects (the decrease of $\Phi $ and the reduction of
structures). One can see in Figures 7 and 6c that transitory peaks and
decays appear around $t=2.7$ for $D_{2},$ $\Phi ,$ $n_{tr}$\ and $s_{i}$.
After that, the potential increases again and reaches a higher level in the
nonlinear regime. The diffusion coefficients $D_{2}$ recovers large values,
but without driving the decay process. The strong diffusion is supported by
the potential, which increases at this stage instead of having a fast decay
as in the previous event. The cause of the different behahiour is the
evolution of the parameters of the spectrum. The dominant wave number $%
k_{2}^{0}$\ (maxima of the spectrum) and the width of the spectrum $\delta
k_{i}\cong \sqrt{2\pi }/\lambda _{i}$ decrease as seen in Figure 6b. As a
consequence, the damping effect of the diffusion $k_{i}^{2}D_{i}$\ is weaker
at small wave numbers and the growth rate of the modes remains positive.

The third time interval in the evolution is characterized by the presence of
the zonal flow modes at amplitudes comparable to $\Phi .$\ The drift
turbulence decays and saturates while the zonal flow amplitude increases.
The EC of the drift potential also saturates ($\lambda _{i}$\ and $k_{2}^{0}$%
\ are roughly constant during this interval). The increase of $V_{2}$ shown
in Figure 6a is determined by the growing contribution of the zonal flow
modes.

Zonal flow provide the explanation for the increase of trapping (Figure 6c)
that seems paradoxical in the condition of the decay of $K_{\ast }^{^{\prime
}}$ below the nonlinearity limit. The structure of the potential contour
lines is modified by the zonal flow potential, which determines larger
structures that capture the ions which are not micro-confined by the drift
turbulence. The fraction of trapped ions and the size of the structures
increase in correlation with the increase of $\Phi _{zf}$ (Figures 6c and
7). A deformation of the trajectory structures appears due to the stronger
increase of $s_{y}$\ that of $s_{x}.$\ They are initially elongated in the
gradient direction and eventually they acquire a shape dominated by the
zonal flow modes. The fraction of trapped ions strongly increases and, at $%
t\cong 7,$ it overcomes the value $n_{tr}=1/2$ ($n=1)$ that corresponds to
negative growth rate of the drift modes (\ref{gamnl}) even in the absence of
diffusion.

The diffusion coefficients decrease in this stage due to the decay of the
potential $\Phi .$ The increase of the diffusion $D_{2}$ determined by the
zonal flow contribution discussed in \cite{VS2013} is not observed here
because the negative growth rate effect leads to the transition in the
quasilinear regime and the condition ($\Phi _{zf}=\Phi >\lambda _{1}$) for
this effect is not fulfilled. The decay of the drift turbulence is caused by
the zonal flow modes through the amplification of trapping rather than by an
increased diffusion.

The evolution of the spectrum of the drift turbulence is presented in Figure
8a, which shows the function $S(0,k_{2})$ at time intervals $\Delta t=1.$\
The spectrum becomes narrower as time increases and its maxima move to
smaller values of $k_{2}.$\ 

The Eulerian correlation of the potential is shown in Figure 8b. The
oscillatory dependence on $x_{2}$ is determined by the narrow peaks of the
spectrum. It determines the small diffusion along the density gradient $%
D_{1}<D_{2}$\ despite the amplitude of the stochastic velocity that is much
larger in this direction ($V_{1}>V_{2}).$\ 

The spectrum of the zonal flow modes $S_{zf}(k_{1})$ is shown in Figure 8c
at the same time moments as $S(0,k_{2}).$\ The two spectra have similar
shapes, with the difference that the zonal flow modes have small tails at
large $k_{1}$ while a very strong cut appears for the drift modes at large $%
k_{2}$. This is the consequence of the diffusive damping that is weak in the
case of the zonal flow modes, as seen in Eq. (\ref{gamzf}).\ The dominant
wave number of the zonal flow modes is around $k_{zf}\cong 1,$ and it has a
weak decrease with time (as also seen in Figure 6b, black curve).\ 

\section{Summary and conclusions}

Particle trajectories in 2-dimensional incompressible velocity fields can
include both random and coherent aspects, which appear as random sequences
of large jumps and trapping or eddying events. This complex behaviour is
determined by the Hamiltonian structure of the equation of motion, and it
effectively appears when the stochastic potential is only weakly perturbed.
The trapped trajectories have a high degree of coherence and they form
intermittent quasi-coherent trajectory structures. Their size and life-time
depend on the characteristic of the turbulence. The distribution of the
displacements is non-Gaussian in the presence of these vortical structures.
We have developed semi-analytical statistical methods for the study of such
complex trajectories, the DTM \cite{V98} and the NSA \cite{VS04}. It was
shown that these methods provide very clear physical images on the nonlinear
transport processes and reasonably good quantitative results.

The quasi-coherent structures, which actually produce a process of
micro-confinement, have strong effects on the transport. The transport is
determined by the long jumps, while the structures represent a diffusion
reservoir, which leads to significant increase of the diffusion coefficients
when the particles are released by a decorrelation mechanism. The transport
process is completely different in the presence of structures in the sense
that the dependence on the parameters is different. \ \ \ 

The quasi-coherent structures have strong effect on the test modes on
turbulent plasmas. This conclusion is drawn from an analytical study of the
drift turbulence that includes the coherent aspects of the trajectories
beside the random ones \cite{Vlad2013}. The frequencies and the growth rates
of the test modes on strongly turbulent plasmas could be determined by the
renormalized propagator procedure. A different perspective on important
aspects of the physics of drift type turbulence in the strongly nonlinear
regime is deduced. The main role in the nonlinear processes is shown to be
played by the ion trajectory structures. The nonlinear damping of drift
modes and the generation of zonal flow modes appear when trapping is strong
due to the ion flows generated by the drift of the potential with the
diamagnetic velocity. The conclusion drawn from this test mode study is that
there is no causality relation between these processes, although there is
time correlation between them. The predator-prey paradigm is not sustained
by these results.

These test particle and test mode studies determine the transport
coefficients and the characteristics of the test modes for given statistical
description of the turbulence, as functions of turbulence parameters. The
conclusions on turbulence and transport evolution base on the analysis of
these results are speculative, because they only suppose that the positive
growth rates of the test modes eventually drive the turbulence in the
nonlinear regime.

We show here that test particle and test mode studies can be connected in an
iterated self-consistent (ISC) theoretical description of turbulence
evolution. It is an approximate Lagrangian method that takes into account
the quasi-coherent aspects of particle trajectories in turbulence. A
computer code based on the ISC was developed for the drift turbulence in
uniform magnetic field. The advantage of this approach is the clear physical
image that can be deduced for complicated nonlinear processes. They are
related to the basic problem of particle advection in turbulence.

The first results show that drift turbulence evolution can have different
characteristics, which essentially depend on the drive parameter.

In the case of weak drive of the drift turbulence, the evolution is
determined by the nonlinear effects of the ion diffusion. They consist not
only of the diffusive damping that reduces the growth rate of the modes, but
also of the modification of the spectrum. The dominant wave number decreases
and the correlation length increase. In particular the increase of $\lambda
_{1}$ leads to values that are larger than those of the normalized potential 
$K_{\ast }^{^{\prime }}$ during the evolution, which shows that trajectory
trapping is negligible. The ion trajectories are random in this case, with
negligible structures.

In the case of strong drive, both random and the quasi-coherent
characteristics of the trajectories are generated during the evolution. The
structures of trajectories influence the evolution of the drift turbulence
by three mechanisms. All of them are connected to the quasi-coherent
trajectory structures. One in the dependence of the growth rates of the test
modes (\ref{gamnl}) on the parameters of the structures (the fraction of
trapped trajectories $n$ and the size of the structures). The increase of $n$
and of $s_{i}$\ determine the decrease of \ $\overline{\gamma }_{d},$\ which
become negative for all wave numbers at $n=1$\ even for $D_{i}=0.$\ The
increase of the structures determines the generation of the zonal flow modes
(as $\overline{\gamma }_{zf}\approx n_{tr}$\ (\ref{gamzf})), which represent
the second possible influence on the drift turbulence. They action is
indirect and consists of the modification of the diffusion and/or of the
structures. The third mechanism is provided by the diffusion, which can be
strongly changed by the structures.

The results presented in the Figures 6-8 show a strong action of the third
mechanism at the beginning of the nonlinear stage, around $t=3.$ Trajectory
trapping determines a strong increase of the diffusion coefficient $D_{2},$
which drives the turbulence back in the quasilinear regime. The process last
for a short time interval, and it is eliminated by the decrease of the wave
numbers of the drift turbulence, which reduces strongly the diffusive
damping in the growth rate (\ref{gamnl}). The decrease of the wave numbers
is due to the narrowing of the spectrum as the effect of increased
diffusion. Thus, a complex transient process of interaction appears at this
stage, which is the result of increased diffusion which determines both the
increase and the decrease of the diffusive damping. The other two mechanisms
determine the evolution at later time when the zonal flow modes reach
amplitudes that are comparable to $\Phi .$ The zonal flow modes determine
the increase of the structures due the the modification of the potential.
The fraction of trapped trajectories increases, the structure have larger
size and they become elongated in the direction of the zona flow modes. This
leads to the decay of the drift turbulence which acquire negative growth
rates. The zonal flow effect is produced through the quasi-coherent
structures, rather than through the diffusion in the system studied here.

In conclusion, the results obtained in this paper bring a new image on
nonlinear processes associated to the generation of quasi-coherence in
particle trajectories. They apply to the drift turbulence in a simple
confinement geometry (uniform magnetic field), but we expect that similar
processes can be found in more realistic models of turbulence. A
semi-analytical method that could be extended to other types of turbulence
was developed.

\bigskip

\textbf{Acknowledgements}

This work was supported by the Romanian Ministry of National Education under
the contract 1EU-10 that is included the Programme of Complementary Research
in Fusion. The views presented here do not necessarily represent those of
the European Commission. The last part of this work was supported on the
contract PN-16470101.

\end{document}